\documentclass[aps,prl,twocolumn,superscriptaddress,floats]{revtex4}
\usepackage{graphicx}
\usepackage{bm}
\usepackage{epsfig}
\input epsf
\usepackage{color,hyperref}
\definecolor{blue}{rgb}{0,0.0,0.9}
\definecolor{green}{rgb}{0.0,0.9,0.0}
\begin{document}
\newcommand{\nsl}[1]{\rlap{\hbox{$\mskip 1 mu /$}}#1}
\newcommand{\sgn}{\operatorname{sgn}}
\preprint{APS/123-QED}
\title{Macroscopic magnetic frustration}
\author{Paula Mellado}
\affiliation{School of Engineering and Applied Sciences, Harvard
University, Cambridge, MA 02138}
\author{Andres Concha}
\affiliation{School of Engineering and Applied Sciences, Harvard
University, Cambridge, MA 02138} \author{L. Mahadevan}
\affiliation{School of Engineering and Applied Sciences, Harvard
University, Cambridge, MA 02138} \affiliation{Department of Physics,
Harvard University, Cambridge, MA 02138}
\begin{abstract}
Although geometrical frustration transcends scale, it has primarily been evoked in  the micro and mesoscopic realm to characterize such phases as spin-ice liquids and glasses and to explain the behavior of such materials as multiferroics, high temperature superconductors, colloids and copolymers. Here we introduce a system of macroscopic ferromagnetic rotors arranged in a planar lattice capable of out-of-plane movement that exhibit the characteristic honeycomb spin ice rules studied and seen so far only in its mesoscopic manifestation. We find that a polarized initial state of this system settles into the honeycomb spin ice phase with relaxation on multiple time scales. We explain this relaxation process using a minimal classical mechanical model which includes Coulombic interactions between magnetic charges located at the ends of the magnets and viscous dissipation at the hinges. Our study shows how macroscopic frustration arises in a purely classical setting that is  amenable to experiment, easy manipulation, theory and computation, and shows phenomena that are not visible in their microscopic counterparts.

\end{abstract}
\maketitle 
Frustration in physical systems commonly arises because geometrical or topological  constraints prevent  global energy minima from being realized.  Although not limited to microscopic phenomena, it is commonly seen in compounds with spins forming lattices with a triangular motif \cite{diep:2004}. In such systems, frustration may lead to the existence of ice selection rules \cite{petrenko.book}  which have been observed  in a variety of materials where spins form networks such as the corner-sharing tetrahedra, known as the Pyrochlore lattice \cite{nature.399.333,gingras.review.2009,RevModPhys.82.53}, leading to monopole-like excitations \cite {castelnovoNATURE2008} and other exotic phases of matter \cite{powell}.  Even though, artificial spin ices \cite{tanPhysRevB2006,qiPRB2008,wangNATURE2006} have shown that frustration can be mimicked by classical magnets, these systems do not account quantitatively for the effects of  inertia, dissipation \cite{lub,parkin,nature_chaikin}, dilution and geometrical disorder because of the mesoscopic scale and fast dynamics of the domain walls ($ \sim 10$ ns) that hinder the understanding of collective dynamics processes. Here we aim to circumvent this situation by introducing a new macroscopic realization of a frustrated magnetic system created using single out-of-plane rotational degree of freedom magnetic rotors, arranged in a kagome lattice, a pattern of corner-sharing triangular plaquettes that dynamically evolves into a spin-ice phase after a magnetic quench.  The ice  phase is reached due to the delicate interplay between inertia, friction and Coulomb-like interactions between the macroscopic magnetic rods. Our prototypical frustrated system has a few advantages  for research in frustrated magnetic systems associated with the ability to (i) tune the interactions through changes in distance and/or orientation between magnets and (ii) examine the lattice relaxation dynamics by direct visualization at a single particle level.  

\begin{figure*}
\begin{center}
\begin{tabular}{cccccc}
\resizebox{62mm}{!}{\includegraphics{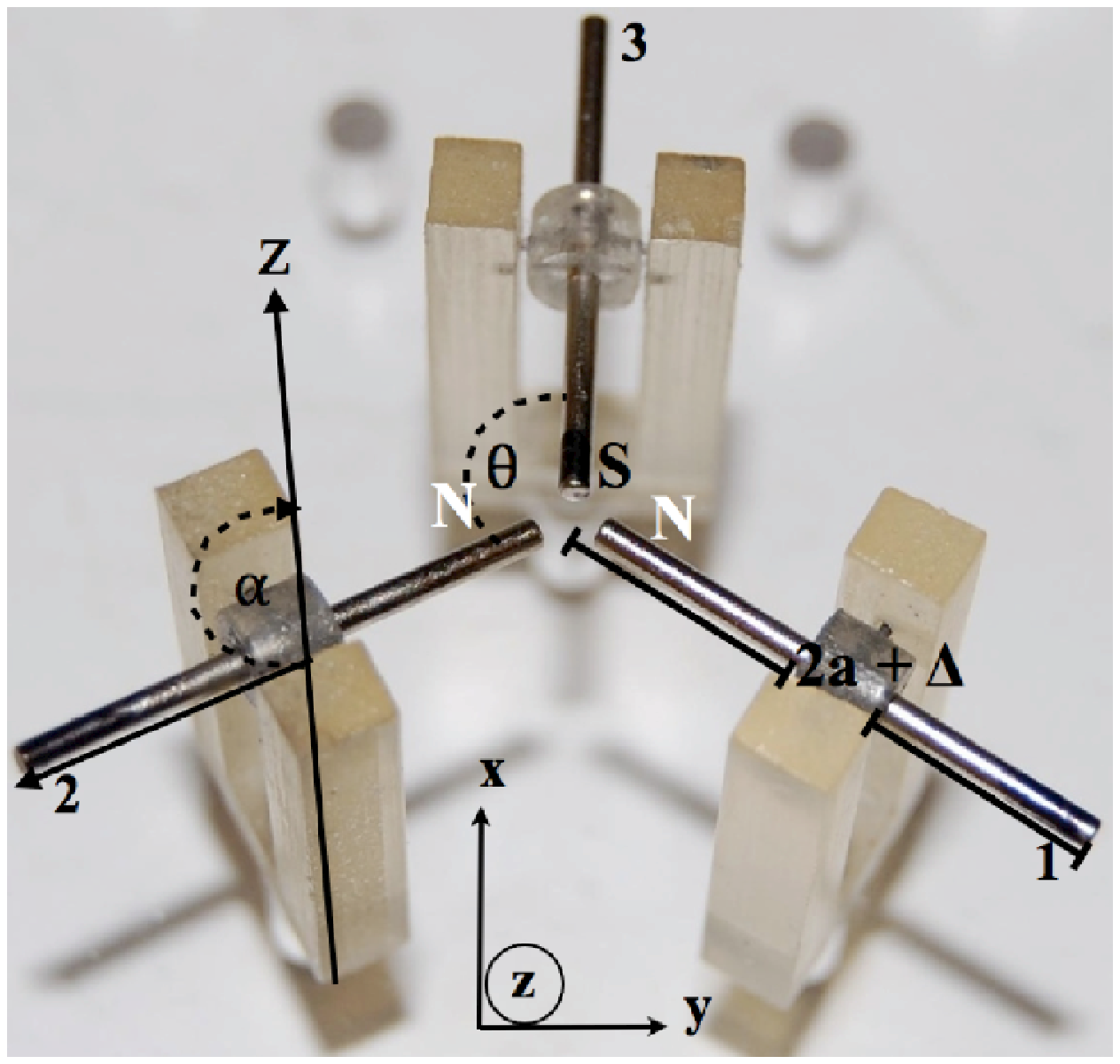}}&
\resizebox{65mm}{!}{\includegraphics{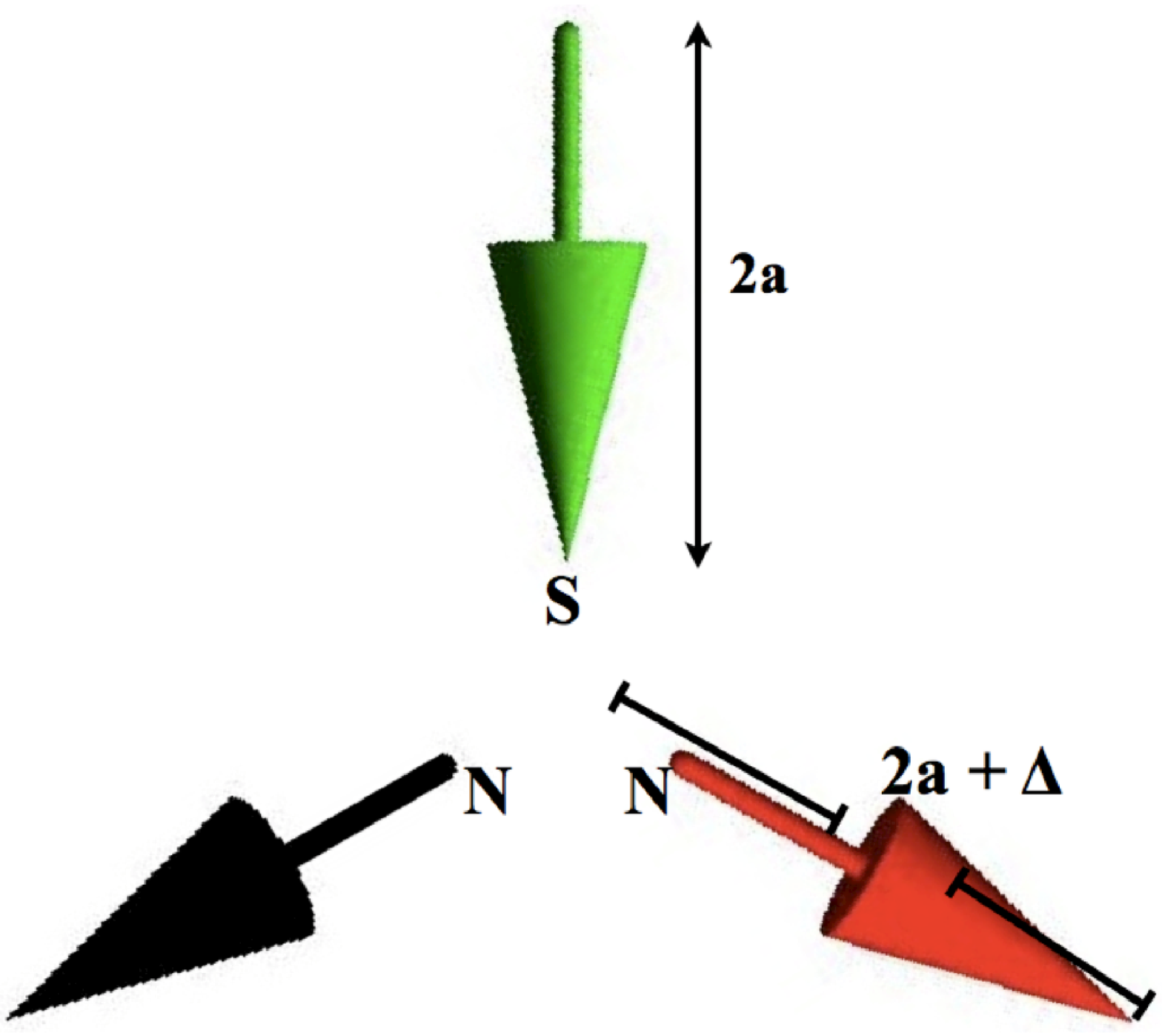}}\\
\resizebox{89mm}{!}{\includegraphics{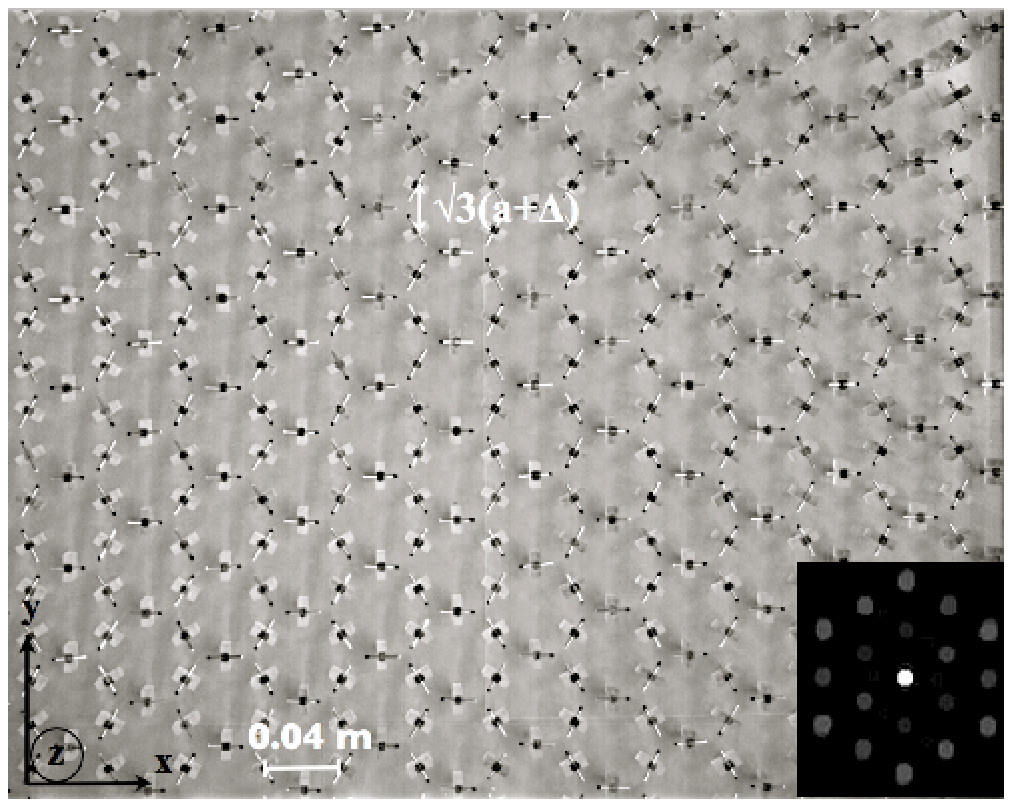}}&
\resizebox{79mm}{!}{\includegraphics{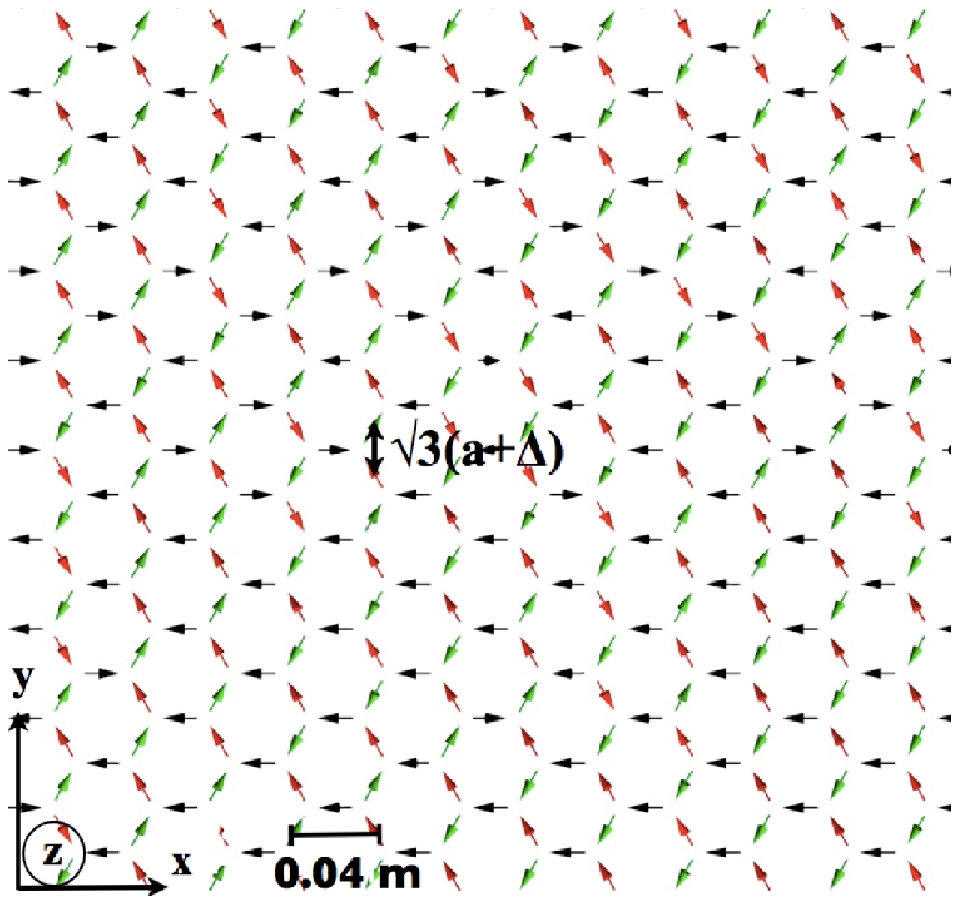}}&
\end{tabular}
\resizebox{75mm}{!}{\includegraphics{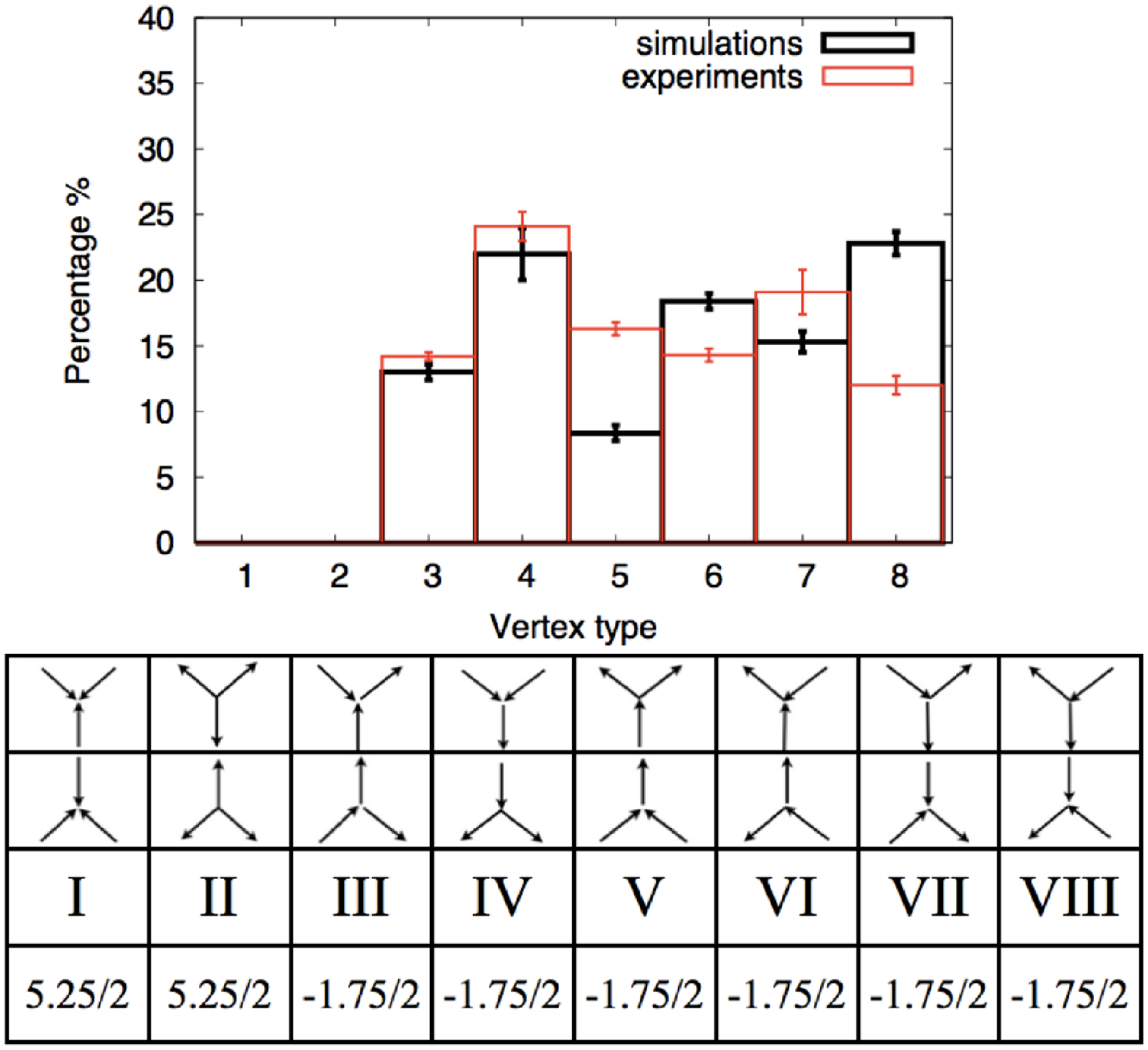}}\\
\end{center}
\caption{\label{FIG1} \textbf{The unit cell star configuration and the honeycomb lattice of magnetic rotors in the spin ice manifold.} $\mathbf{a,}$  A triad of magnetic rotors (lying in one of the sublattices indicated as 1, 2 and 3) having length $L=2a=1.9\times 10^{-2}$ m, diameter $d=1.5\times 10^{-3}$ m, mass $M=0.28 \times10^{-3}$ Kg and saturation magnetization $M_{s}= 1.2 \times 10^{6}$ A $m^{-1}$ are located at $\theta=120^{\circ}$ respect to each other. The out of plane degree of freedom is denoted by the polar angle $\alpha$. Painted in black the magnet south pole (S) is distinguished from its north pole (N). $\mathbf{b,}$ The numerical equivalent magnetic triad having the same experimental parameters. In this case the point of the arrow denotes the S pole. $\mathbf{c,}$ Picture of the lattice with its centers located at distance $l=\sqrt{3}(a+\Delta)$ with the n rods lying in the $x-y$ plane fulfilling the honeycomb spin ice rules. Inset shows the Fourier transform of the lattice. $\mathbf{d,}$ The numerical equivalent lattice with same parameters as the experimental one. $\mathbf{e,}$ Up: Histograms taken from 10 experiments and simulations showing the experimental (red) and numerical (black) distribution of vertices. Bottom:  Table with the local energy of the eight vertex configurations possible in the honeycomb lattice, in units of $D=10^{-5}$ J (Supplementary S6).}
\end{figure*}

A minimal macroscopic realization of local frustration  can be seen easily in a $120^{\circ}$ star configuration using three ferromagnetic rods with their hinges on a plane (Fig. \ref{FIG1}a). The rods have length $L=2a=1.9\times 10^{-2}$ m, diameter $d=1.5\times 10^{-3}$ m, mass $M=0.28 \times 10^{-3}$ Kg and saturation magnetization $M_{s}= 1.2 \times 10^{6}$ A $m^{-1}$. By design the only allowed motions for the rotors are rotations in the polar direction $\alpha$.  The hinges supporting the rods were placed at the sites of a kagome lattice with lattice constant $l=\sqrt{3}(a+\Delta)$ where $\Delta$ is the shortest distance between the tips and the nearest vertex center and $\Delta/L\sim 0.2$ (Fig. \ref{FIG1}a), so that when in the $x-y$ plane, the magnets realize the bonds of a honeycomb lattice. The magnetization of a rotor $i$ is defined as the vector $\mathbf{m_{i}}$ joining its N to its S pole, thus $\mathbf{m_{i}}$ is the coarse-grained spin variable for each magnet.  When all three magnets are close to each other, the lowest energy configuration consists of one pole being different from the others, leading to a frustrated state consisting of permutations of NNS or SSN (S=south pole, N = north pole)  that correspond to the honeycomb spin ice rules \cite{qiPRB2008,melladoPRL2010}.  With this unit-cell plaquette, we prepare a polarized lattice of $n=352$ of these magnetic rotors, with an unavoidable geometrical disorder in the azimuthal orientation of the rotors, $\theta$,  due to lattice imperfections $\delta \theta_{\rm{max}}\sim 2^{o}$; this  follows a Gaussian distribution with mean $\bar{\delta\theta}=1.2^{\circ}$. We oriented the S poles of all rotors out of the plane by applying a strong magnetic field along the $\hat{z}$ direction $B_{z}=3.2\times 10^{-3}$ T (Supplementary Information S4). At $t=0$  the field was switched off, to allow for the lattice to relax, a procedure that was repeated several times. After about 2  seconds, all the rotors had reached equilibrium configurations very near the $x-y$ plane (the non-planarity out of the $x-y$ plane $\delta\alpha \sim 10^{\circ}$ on average) and in the honeycomb spin ice manifold. Figure  \ref{FIG1}e, shows a picture of the lattice where all the rods fulfill the ice rules. The experimental distribution of vertices is shown in the (red) bars of Figure \ref{FIG1}c. We find all vertices falling into the six low energy (spin ice) configurations while high energy states (type 1 and 2) are absent. 

This macroscopic spin ice consists of elemental rotor units that constitute a frustrated triad which we characterize at a static and a dynamic level (Supplementary Information S1, S2 and S3). This allows us to use a dipolar dumbbell approach to the magnets \cite{castelnovoNATURE2008}, determine the charge $q=\pi M_{s}d^{2}/4\sim 2.03$ A m, at each pole, find the damping time scale for an isolated rod $\tau_{D}\sim 1$ s  and examine how Coulomb interactions and geometrical disorder in $\theta$ and $\Delta$ control the orientations of the rods relative to each other.  On a collective level,  the relaxation of the lattice from the $\hat{z}$ polarized state to the spin ice manifold may be characterized in terms of the correlation between nearest neighbor spins $\alpha$ and $\beta$, with $S_{\alpha }S_{\beta}=1$ when $\mathbf{m_{\alpha}}\cdot \mathbf{m_{\beta}}$ is positive, $S_{\alpha}S_{ \beta}=-1$ otherwise. From high speed movies (400 fps), we extracted the full time trajectory $\alpha_{i}(t)$ of the $i-th$ rotor (Supplementary S4 and Movie MS1) and computed the spin-spin correlations.  

\begin{figure*}
\begin{center}
\begin{tabular}{cccccc}
\resizebox{60mm}{!}{\includegraphics{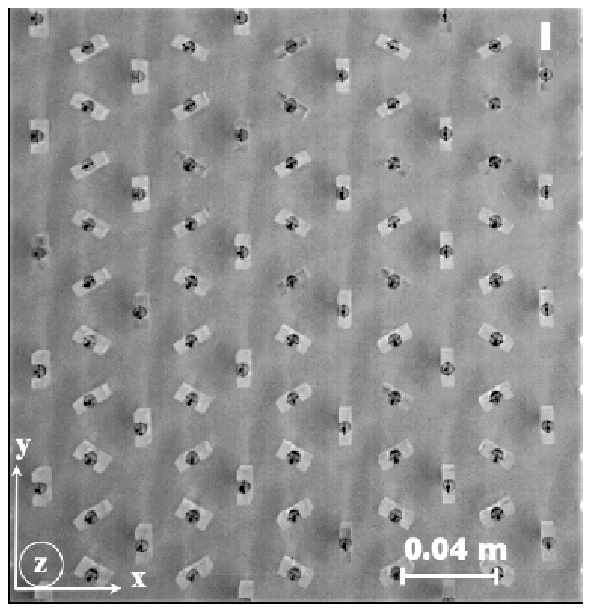}}&
\resizebox{60mm}{!}{\includegraphics{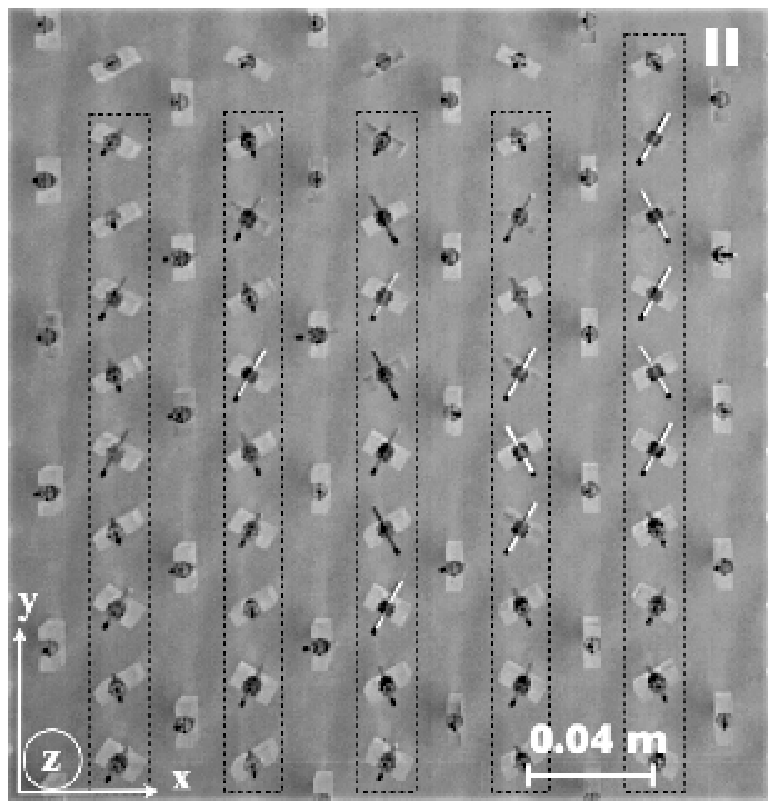}}&
\resizebox{60mm}{!}{\includegraphics{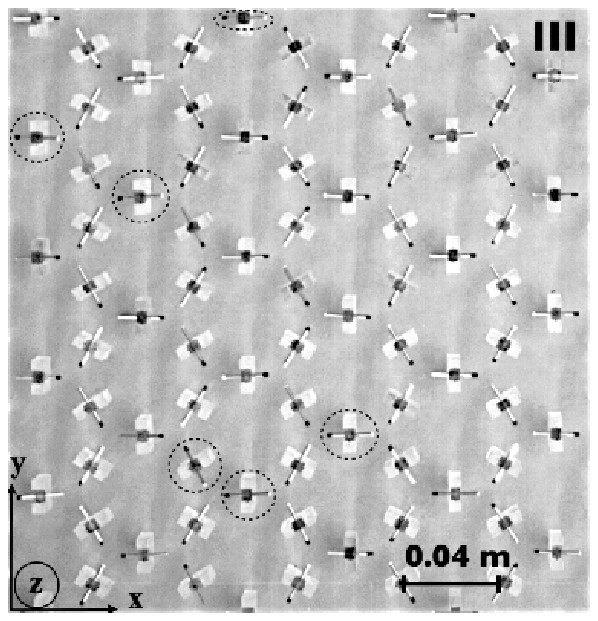}}&
\end{tabular}\\
\resizebox{89mm}{!}{\includegraphics{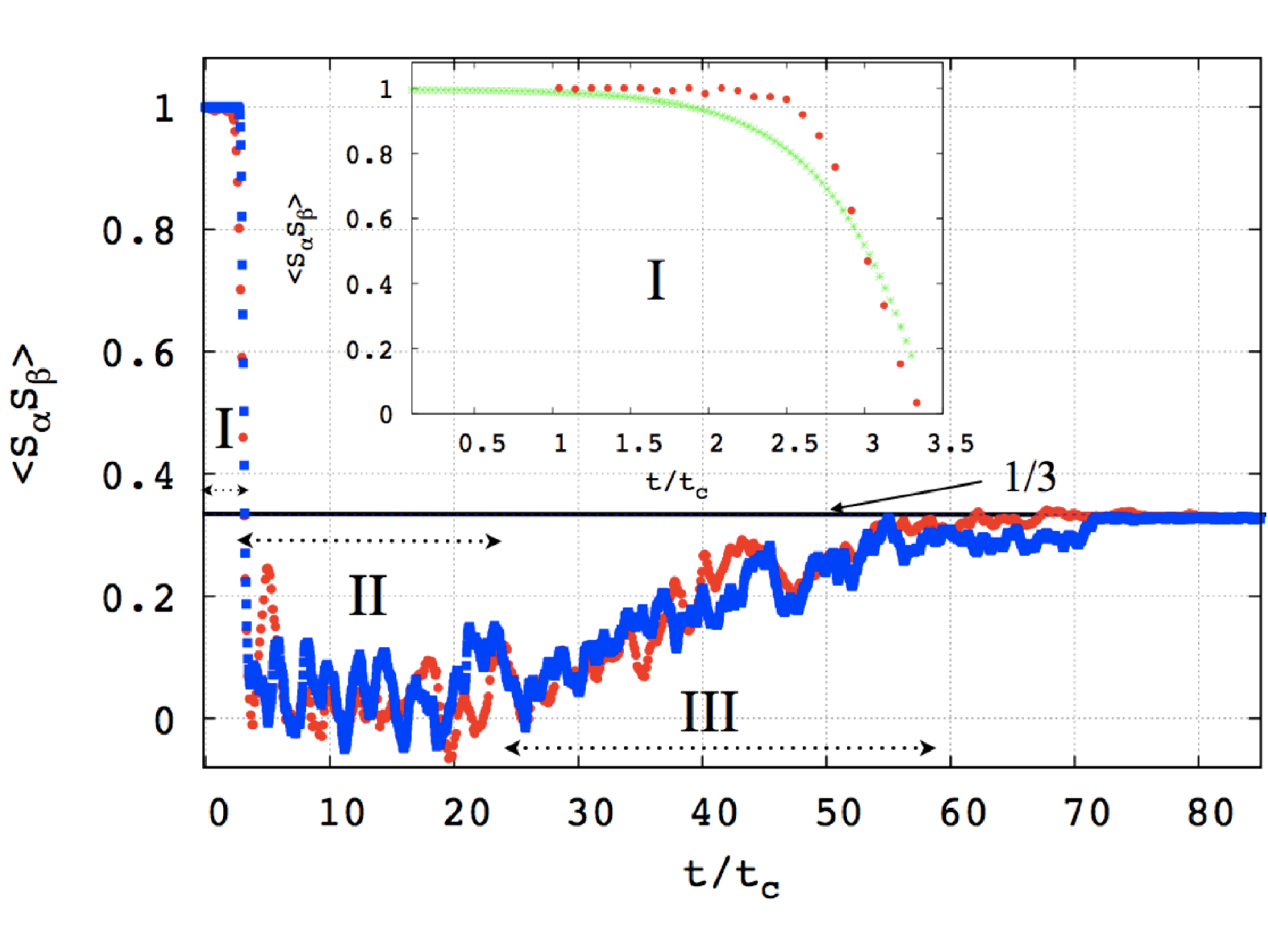}}\\
\end{center}
\caption{\label{FIG2}\textbf{Lattice dynamics characterized by nearest neighbor spin correlations, $\langle S_{\alpha}S_{\beta}\rangle$.} $\mathbf{a,}$ Stage I: once $B_{z}$ is turned off, the rotors originally pointing along $\hat{z}$ break their axial symetry. $\mathbf{b,}$ with the image showing the end of stage I and the onset of stage II when rods rotate respect to their center of mass yielding a plateau in $\langle S_{\alpha}S_{\beta}\rangle$. $\mathbf{c,}$ with a snapshot of the rods oscillating in stage III. $\mathbf{d,}$ In red experimental data obtained via image processing, in blue molecular dynamics simulation results from the numerical solution of equation (S4) where the full coulomb contributions from all neighbors is taken into account. At t=0 all S poles point along $\hat{z}$. Stage I is dominated by Coulomb interactions between rods and characterized by the Coulomb time scale $t_{c}$.  In Stage II, all rotors spin  until dissipation damps out the spin in favor of oscillations, leading to Stage III where they exhibit damped oscillations. After relaxation the rods lie in the x-y plane in a honeycomb spin ice magnetic configuration, with its characteristic nearest neighbor spin correlations $\langle S_{\alpha}S_{\beta}\rangle=1/3$ (solid line). Inset:  Experimentally measured value of $\langle S_{\alpha}S_{\beta}\rangle$ during the initial explosive evolution (red) compared with $\cos(\alpha)$ where $\alpha$ is the solution of Supplementary equation (S2) for one rotor interacting with two neighbors, in the absence of damping and external torques (green).}
\end{figure*}

We find that there are three stages in the spin relaxation process. In stage I, corresponding to the first $\sim 0.07$ s, the rotors break their initial axial symmetry, Figure \ref{FIG2}a, and correlations decay rapidly with a characteristic Coulomb time scale  $t_{c} \sim 0.02$ s, Figure \ref{FIG2}d, which is the shortest time scale in the lattice relaxation, with $t_{c}\sim\frac{\sqrt{a I/\mu_{0}}}{q}$  dominated by internal Coulomb interactions for the relaxation of a rotor interacting with two neighbors (Supplementary S4) in the absence of damping and external torques (inset of Figure \ref{FIG2}d).  Next,  magnets of sub-lattices 1 and 2 (Fig.\ref{FIG1}a) organize in head to tail chains along the $\hat{y}$ direction, while those belonging to sub-lattice 3 still remain non-planar, Figure \ref{FIG2}b. In Stage II, once the sub-lattice 3 becomes planar, all the rods spin continuously leading to a plateau in the spin correlations (Fig.\ref{FIG2}d); eventually the kinetic energy of the rotors has been dissipated sufficiently that the rotors oscillate rather than spin.  For our experimental parameters (Fig.\ref{FIG1}a), the phase space trajectory changes from librations to damped oscillations  after $0.45$ s (Supplementary Fig. S7); the rotors typically average about four full rotations before they switch to oscillations. Finally, in Stage III (Fig.\ref{FIG2}c  and Fig.\ref{FIG2}d) the rods oscillate without full rotations: when we fit the experimental dynamics at this state to a decaying exponential we find  $t_{d} \sim \tau_{D}$, thus this stage is dominated by dissipative effects. 

To understand these different dynamical regimes, we  performed molecular dynamics simulations of the massive underdamped rotors interacting through the full long-range internal Coulomb interactions between all the rods in the lattice using a Verlet algorithm (Supplementary S5, Fig.S10 and Movie SM2). In Figure \ref{FIG2}d, we see that the computed nearest neighbor spin correlations for the relaxation of the numerical lattice has the same three qualitative different regimes as in the experiments when the lattice relaxes from a polarized state to its spin ice manifold. Furthermore, the Coulomb and damping time scales for stage I and III as well as the plateau featuring stage II are in good agreement with experiments. The observed high frequency fluctuations in $\langle S_{\alpha}S_{\beta}\rangle (t)$ in both, experiments and simulations, are due to the Coulomb coupling between rods which rapidly reorient while they relax due to the fluctuations in the internal magnetic field.

Having examined the dynamics of relaxation to the spin ice state, we now turn to the lattice response when a dipole with charge $|Q^{e}|$ at each pole and length $L^{e}$, at a vertical distance $h$ underneath the relaxed lattice is moved along one of the three sub-lattices at speed $v$  (Supplementary Fig.S9). For an isolated rotor, the critical torque that is required to destabilize the planar configuration is given by $\mathit{T_{c}}\sim 2aB_{c}q$, where $B_{c}$ is the applied magnetic field; experiments on many rotors yielded an average  $B_{c}\sim (2.4 \pm 0.1)\times 10^{-4}$ T. Equivalently, the threshold distance at which the external field will overcome both internal Coulomb interactions and static friction  is given by $h^{*}\sim \sqrt{Q^{e}qa/\mathit{T_{c}}}$. Dynamically, the internal Coulomb interactions set a time scale for small out-of-plane oscillations of the rotors in the lattice, given by  $\tau_{ph}\sim \sqrt{\Delta^{2}I/\mu_{0}q^{2}a} \sim 0.01 s$ for the experimental parameters at hand. Thus, there are two dimensionless quantities that determine the response to the external perturbation: the ratio between phononic and kinetic time scales $v\tau_{ph}/a$ and the ratio between internal and external magnetic forces, $F^{int}/F^{ext}=qh^{2}/(Q^{e}\Delta^{2})$. 

\begin{figure*}
\begin{center}
\begin{tabular}{cccccc}
\resizebox{89mm}{!}{\includegraphics{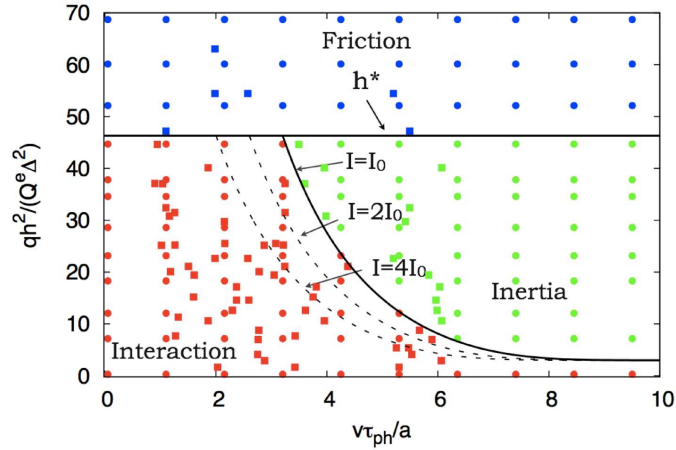}}\\
\end{tabular}
\end{center}
\caption{\label{FIG3} \textbf{Phase diagram of the lattice dynamical response to an external perturbation.} The horizontal axis shows the dimensionless ratio of the kinetic and phononic time scales with $v$ the speed of an external dipole, while the vertical axis shows the dimensionless ratio of the internal and the external magnetic forces due to an external dipole of strength $Q$ located at distance $h$ from the lattice (see text for details). Experimental and numerical data shown in squares and circles respectively, and colors define the nature of the lattice dynamical response to the external perturbation. We see that the dynamics may be broken up into a frictionally dominated, interaction dominated or inertially dominated regime as a function of the relative magnitude and rate of external forcing. }
\end{figure*}

In Figure \ref{FIG3}, we characterize the phase diagram of the dynamical response of the spin-ice lattice in terms of these dimensionless parameters.  For $h<h^{*}$, the lattice is disturbed only in a band of width $D(h)\sim \sqrt{(h^{*2} h)^{2/3}-h^{2}}$ centered along the trajectory of the moving external dipole, based only on local interactions, static friction, and interactions with the external dipole (Supplementary Fig.S9). For large $v$,  $\tau_{ph}/\tau_k \gg 1$ so that the rotors have little time to respond and barely oscillate in an inertia-dominated regime. In the opposite limit, large amplitude oscillations and flips are apparent as there is enough time for the rotors to interact with the external dipole. Our results for these regimes show that the simulations (filled circles) and experiments (filled squares) agree.  The solid line defines a threshold of the RMS fluctuations for the oscillations of all the rods $\delta_{Li}^{th}=0.5$ (Supplementary S4) separating the regimes. To understand this, we resort to a simple single rod approximation where the impulsive response of a rotor due to a long dipole located at a distance $d(t)=\sqrt{h^{2}+(v t)^{2}}$, balances the change in its angular momentum yielding $h\sim h^{*}(Q^{e}qa/I)^{3/2}/v^{3}$,  consistent with the observations when $v\tau_{ph}/a \gg 1$. Varying inertia from $I_{0}$ to $4I_{0}$ using our simulations we confirmed that as $I$ grows, the boundary between interaction and inertial regime shift to the left; the inertial regime is reached for smaller values of $v\tau_{ph}/a$ and $qh^{2}/(Q^{e}\Delta^{2})$. When $h\gg h^{*}$, the Coulomb force due to the external field is not able to overcome the combined effects of static friction and internal Coulomb interactions, and the lattice falls into a friction dominated one in which oscillations are not apparent. 

Our spin ice phase emerges in a system of damped macroscopic rotors, purely driven by interactions in a classical mechanical setting that differs from those found in its micro and mesoscopic relatives. Using a minimal model we can capture the dynamical evolution of the collection of rotors in the lattice observed in our experiments and reproduce the three main stages of lattice relaxation from a polarized state: explosive behavior lasting $t_c$,  dissipative librations and damped oscillations. The advantages of studying this macroscopic realization beyond the present work include the fact that (i) the interactions can be tuned through changes in the diameter of magnets or distance or orientation between them (Supplementary Fig. S4), (ii) inertial and dissipative effects can be studied by controlling the friction coefficient at the hinges as well as the mass of the rods, (iii) the effect of vacancies or random dilution can be examined by removing rotors from the lattice (iv) the lattice relaxation dynamics can be directly visualized at single particle level and (v) the system can be easily generalized to three dimensions (3-D) by stacking plates with hinged rotors along the $z$ direction. Indeed a minimal 3-D realization is shown in Supplementary Fig. S12: a tetrahedral configuration like the one found in the Pyrochlore lattice was created placing three acrylic plates one on the top of the other, the bottom and top plates contain three rotors defining an equilateral triangle and the middle plate contains one rotor located equidistant from the others. The ease of fabrication, manipulation and measurement and the study of a variety of soft modes in  artificial lattices in a system that is nearly five orders of magnitude larger and slower than its mesoscopic counterpart suggests that there is a new class of phenomena waiting to be explored in macroscopic frustrated systems.

\begin{thebibliography}{14}
\expandafter\ifx\csname natexlab\endcsname\relax\def\natexlab#1{#1}\fi
\expandafter\ifx\csname bibnamefont\endcsname\relax
  \def\bibnamefont#1{#1}\fi
\expandafter\ifx\csname bibfnamefont\endcsname\relax
  \def\bibfnamefont#1{#1}\fi
\expandafter\ifx\csname citenamefont\endcsname\relax
  \def\citenamefont#1{#1}\fi
\expandafter\ifx\csname url\endcsname\relax
  \def\url#1{\texttt{#1}}\fi
\expandafter\ifx\csname urlprefix\endcsname\relax\def\urlprefix{URL }\fi
\providecommand{\bibinfo}[2]{#2}
\providecommand{\eprint}[2][]{\url{#2}}

\bibitem[{\citenamefont{Diep}(2004)}]{diep:2004}
\bibinfo{editor}{\bibfnamefont{H.~T.} \bibnamefont{Diep}}, ed.,
  \emph{\bibinfo{title}{Frustrated Spin Systems}} (\bibinfo{publisher}{World
  Scientific}, \bibinfo{year}{2004}).

\bibitem[{\citenamefont{Petrenko and Whitworth}(1999)}]{petrenko.book}
\bibinfo{author}{\bibfnamefont{V.~F.} \bibnamefont{Petrenko}} \bibnamefont{and}
  \bibinfo{author}{\bibfnamefont{R.~W.} \bibnamefont{Whitworth}},
  \emph{\bibinfo{title}{Physics of Ice}} (\bibinfo{publisher}{Oxford University
  Press}, \bibinfo{year}{1999}).

\bibitem[{\citenamefont{Ramirez et~al.}(1999)\citenamefont{Ramirez, Hayashi,
  Cava, Siddharthan, and Shastry}}]{nature.399.333}
\bibinfo{author}{\bibfnamefont{A.~P.} \bibnamefont{Ramirez}},
  \bibinfo{author}{\bibfnamefont{A.}~\bibnamefont{Hayashi}},
  \bibinfo{author}{\bibfnamefont{R.~J.} \bibnamefont{Cava}},
  \bibinfo{author}{\bibfnamefont{R.}~\bibnamefont{Siddharthan}},
  \bibnamefont{and} \bibinfo{author}{\bibfnamefont{B.~S.}
  \bibnamefont{Shastry}}, \bibinfo{journal}{Nature}
  \textbf{\bibinfo{volume}{399}}, \bibinfo{pages}{333} (\bibinfo{year}{1999}).

\bibitem[{\citenamefont{Gingras}(2011)}]{gingras.review.2009}
\bibinfo{author}{\bibfnamefont{M.~J.~P.} \bibnamefont{Gingras}}, in
  \emph{\bibinfo{booktitle}{Introduction to Frustrated Magnetism}}, edited by
  \bibinfo{editor}{\bibfnamefont{C.}~\bibnamefont{Lacroix}},
  \bibinfo{editor}{\bibfnamefont{P.}~\bibnamefont{Mendels}}, \bibnamefont{and}
  \bibinfo{editor}{\bibfnamefont{F.}~\bibnamefont{Mila}}
  (\bibinfo{publisher}{Springer Verlag}, \bibinfo{year}{2011}).

\bibitem[{\citenamefont{Gardner et~al.}(2010)\citenamefont{Gardner, Gingras,
  and Greedan}}]{RevModPhys.82.53}
\bibinfo{author}{\bibfnamefont{J.~S.} \bibnamefont{Gardner}},
  \bibinfo{author}{\bibfnamefont{M.~J.~P.} \bibnamefont{Gingras}},
  \bibnamefont{and} \bibinfo{author}{\bibfnamefont{J.~E.}
  \bibnamefont{Greedan}}, \bibinfo{journal}{Rev. Mod. Phys.}
  \textbf{\bibinfo{volume}{82}}, \bibinfo{pages}{53} (\bibinfo{year}{2010}).

\bibitem[{\citenamefont{{Castelnovo} et~al.}(2008)\citenamefont{{Castelnovo},
  {Moessner}, and {Sondhi}}}]{castelnovoNATURE2008}
\bibinfo{author}{\bibfnamefont{C.}~\bibnamefont{{Castelnovo}}},
  \bibinfo{author}{\bibfnamefont{R.}~\bibnamefont{{Moessner}}},
  \bibnamefont{and} \bibinfo{author}{\bibfnamefont{S.~L.}
  \bibnamefont{{Sondhi}}}, \bibinfo{journal}{Nature}
  \textbf{\bibinfo{volume}{451}}, \bibinfo{pages}{42} (\bibinfo{year}{2008}).

\bibitem[{\citenamefont{Powell}(2011)}]{powell}
\bibinfo{author}{\bibfnamefont{S.}~\bibnamefont{Powell}},
  \bibinfo{journal}{Phys. Rev. B} \textbf{\bibinfo{volume}{84}},
  \bibinfo{pages}{094437} (\bibinfo{year}{2011}).

\bibitem[{\citenamefont{{Tanaka} et~al.}(2006)\citenamefont{{Tanaka}, {Saitoh},
  {Miyajima}, {Yamaoka}, and {Iye}}}]{tanPhysRevB2006}
\bibinfo{author}{\bibfnamefont{M.}~\bibnamefont{{Tanaka}}},
  \bibinfo{author}{\bibfnamefont{E.}~\bibnamefont{{Saitoh}}},
  \bibinfo{author}{\bibfnamefont{H.}~\bibnamefont{{Miyajima}}},
  \bibinfo{author}{\bibfnamefont{T.}~\bibnamefont{{Yamaoka}}},
  \bibnamefont{and} \bibinfo{author}{\bibfnamefont{Y.}~\bibnamefont{{Iye}}},
  \bibinfo{journal}{Phys. Rev. B} \textbf{\bibinfo{volume}{73}},
  \bibinfo{eid}{052411} (\bibinfo{year}{2006}).

\bibitem[{\citenamefont{Qi et~al.}(2008)\citenamefont{Qi, Brintlinger, and
  Cumings}}]{qiPRB2008}
\bibinfo{author}{\bibfnamefont{Y.}~\bibnamefont{Qi}},
  \bibinfo{author}{\bibfnamefont{T.}~\bibnamefont{Brintlinger}},
  \bibnamefont{and} \bibinfo{author}{\bibfnamefont{J.}~\bibnamefont{Cumings}},
  \bibinfo{journal}{Phys. Rev. B} \textbf{\bibinfo{volume}{77}},
  \bibinfo{pages}{094418} (\bibinfo{year}{2008}).

\bibitem[{\citenamefont{{Wang} et~al.}(2006)\citenamefont{{Wang}, {Nisoli},
  {Freitas}, {Li}, {McConville}, {Cooley}, {Lund}, {Samarth}, {Leighton},
  {Crespi} et~al.}}]{wangNATURE2006}
\bibinfo{author}{\bibfnamefont{R.~F.} \bibnamefont{{Wang}}},
  \bibinfo{author}{\bibfnamefont{C.}~\bibnamefont{{Nisoli}}},
  \bibinfo{author}{\bibfnamefont{R.~S.} \bibnamefont{{Freitas}}},
  \bibinfo{author}{\bibfnamefont{J.}~\bibnamefont{{Li}}},
  \bibinfo{author}{\bibfnamefont{W.}~\bibnamefont{{McConville}}},
  \bibinfo{author}{\bibfnamefont{B.~J.} \bibnamefont{{Cooley}}},
  \bibinfo{author}{\bibfnamefont{M.~S.} \bibnamefont{{Lund}}},
  \bibinfo{author}{\bibfnamefont{N.}~\bibnamefont{{Samarth}}},
  \bibinfo{author}{\bibfnamefont{C.}~\bibnamefont{{Leighton}}},
  \bibinfo{author}{\bibfnamefont{V.~H.} \bibnamefont{{Crespi}}},
  \bibnamefont{et~al.}, \bibinfo{journal}{Nature}
  \textbf{\bibinfo{volume}{439}}, \bibinfo{pages}{303} (\bibinfo{year}{2006}).

\bibitem[{\citenamefont{{Han} et~al.}(2008)\citenamefont{{Han}, {Shokef},
  {Alsayed}, {Yunker}, {Lubensky}, and {Yodh}}}]{lub}
\bibinfo{author}{\bibfnamefont{Y.}~\bibnamefont{{Han}}},
  \bibinfo{author}{\bibfnamefont{Y.}~\bibnamefont{{Shokef}}},
  \bibinfo{author}{\bibfnamefont{A.~M.} \bibnamefont{{Alsayed}}},
  \bibinfo{author}{\bibfnamefont{P.}~\bibnamefont{{Yunker}}},
  \bibinfo{author}{\bibfnamefont{T.~C.} \bibnamefont{{Lubensky}}},
  \bibnamefont{and} \bibinfo{author}{\bibfnamefont{A.~G.}
  \bibnamefont{{Yodh}}}, \bibinfo{journal}{Nature}
  \textbf{\bibinfo{volume}{456}}, \bibinfo{pages}{898} (\bibinfo{year}{2008}).

\bibitem[{\citenamefont{{Thomas} et~al.}(2010)\citenamefont{{Thomas}, {Moriya},
  {Rettner}, and {Parkin}}}]{parkin}
\bibinfo{author}{\bibfnamefont{L.}~\bibnamefont{{Thomas}}},
  \bibinfo{author}{\bibfnamefont{R.}~\bibnamefont{{Moriya}}},
  \bibinfo{author}{\bibfnamefont{C.}~\bibnamefont{{Rettner}}},
  \bibnamefont{and} \bibinfo{author}{\bibfnamefont{S.~S.~P.}
  \bibnamefont{{Parkin}}}, \bibinfo{journal}{Science}
  \textbf{\bibinfo{volume}{330}}, \bibinfo{pages}{1810} (\bibinfo{year}{2010}).

\bibitem[{\citenamefont{Irvine et~al.}(2010)\citenamefont{Irvine, Vitelli, and
  Chaikin}}]{nature_chaikin}
\bibinfo{author}{\bibfnamefont{W.~T.~M.} \bibnamefont{Irvine}},
  \bibinfo{author}{\bibfnamefont{V.}~\bibnamefont{Vitelli}}, \bibnamefont{and}
  \bibinfo{author}{\bibfnamefont{P.~M.} \bibnamefont{Chaikin}},
  \bibinfo{journal}{Nature} \textbf{\bibinfo{volume}{468}},
  \bibinfo{pages}{947} (\bibinfo{year}{2010}).

\bibitem[{\citenamefont{Mellado et~al.}(2010)\citenamefont{Mellado, Petrova,
  Shen, and Tchernyshyov}}]{melladoPRL2010}
\bibinfo{author}{\bibfnamefont{P.}~\bibnamefont{Mellado}},
  \bibinfo{author}{\bibfnamefont{O.}~\bibnamefont{Petrova}},
  \bibinfo{author}{\bibfnamefont{Y.}~\bibnamefont{Shen}}, \bibnamefont{and}
  \bibinfo{author}{\bibfnamefont{O.}~\bibnamefont{Tchernyshyov}},
  \bibinfo{journal}{Phys. Rev. Lett.} \textbf{\bibinfo{volume}{105}},
  \bibinfo{pages}{187206} (\bibinfo{year}{2010}).

\end{thebibliography}
\end{document}


\newcommand{\nsl}[1]{\rlap{\hbox{$\mskip 1 mu /$}}#1}
\newcommand{\sgn}{\operatorname{sgn}}
\renewcommand{\thefigure}{S\arabic{figure}}
\renewcommand{\theequation}{S\arabic{equation}} 

\title{\bf Supplementary Information for \\``Macroscopic magnetic frustration" \\ by P. Mellado, A. Concha and L. Mahadevan}

\maketitle

\section*{S1. Coulomb like interaction between two magnetic rods}
\begin{figure*}[h]
\centering
\includegraphics[width=0.7\textwidth]{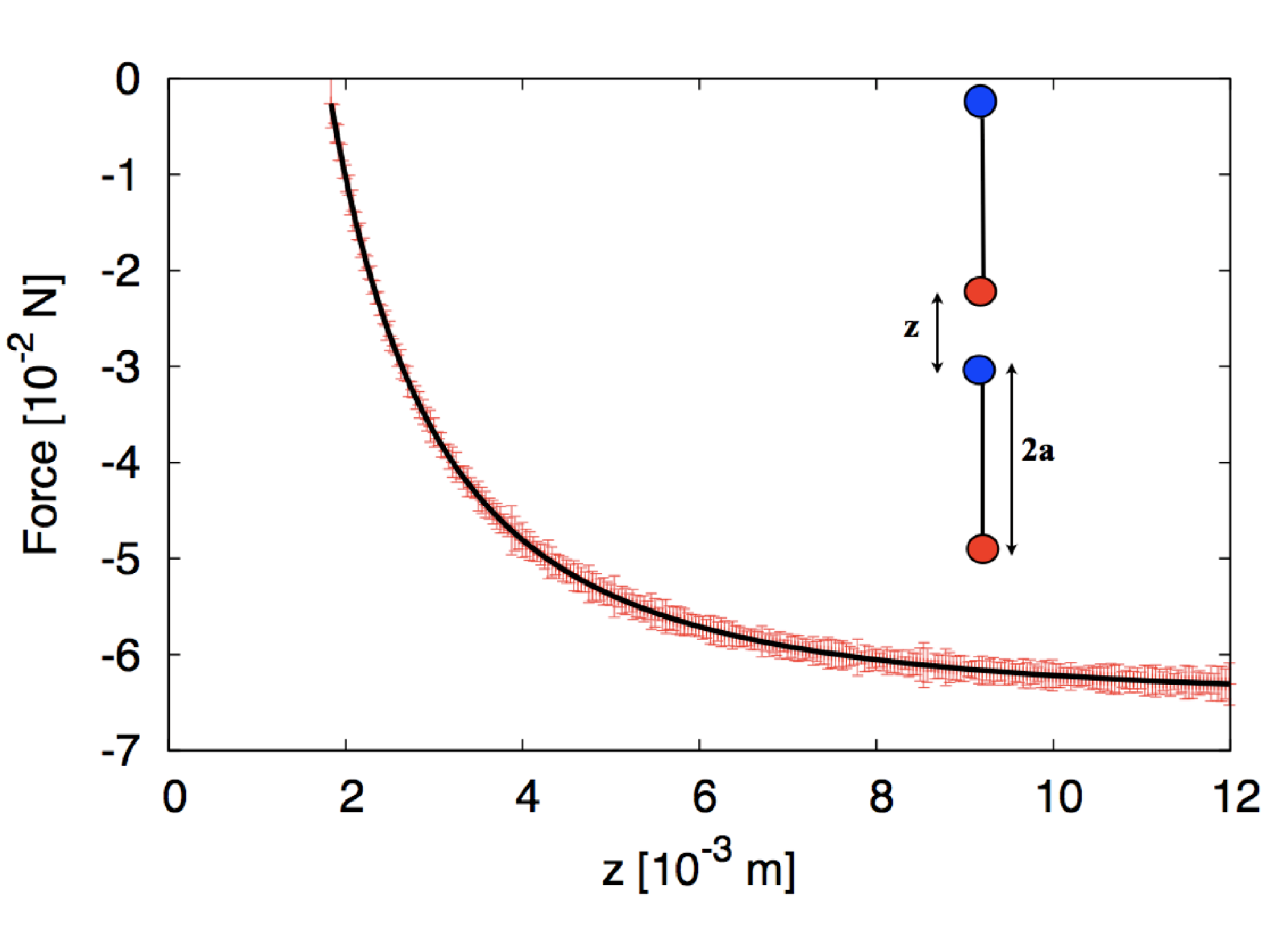}
\caption{\label{FIGS1} \textbf{Force vs. distance between two Neodymium rods.}  In red are shown the experimental data with error bars of the attractive force between  two magnetic  rods, as  a function of  the distance  between the two  closest faces of  the rods. Black  line represents  the  best fit  obtained by  using the Coulomb law (Eq.(\ref{dumbbell})), from where we obtained the magnitude of the magnetic charges $q=2.03 \pm 0.08$ A m. The inset illustrates a typical tensile experiment.}
\end{figure*}

In order to compute the magnetic charge parameter $q$ used in our simulations, we measured the force-distance relation for several bar magnets using an Instron 5544  system to estimate the charge (pole strength) at the end of each bar in the dumbbell model. Two magnetic rods of equal dimensions were aligned with respect to each other at a distance $z$ along the vertical axis and with opposite poles facing each other. Then, one of the rods was pulled apart and the interaction  force measured, with a typical outcome shown in Figure \ref{FIGS1}, where the experimental data for two rods in  the configuration shown in the  inset of Figure \ref{FIGS1} is shown in red. The continuous black line is the fit obtained  from a dumbbell  model, where interactions between magnetic poles follow a Coulomb  law, and the  only free parameter  is the  magnetic charge  at the  tips of  the  rod:
\begin{eqnarray}
F(z)=\frac{\mu_{0}q^{2}}{4\pi}\left(\frac{1}{z^2}-\frac{2}{(z+2a)^{2}}+\frac{1}{(z+4a)^{2}}\right)
\label{dumbbell}
\end{eqnarray} 

From the fit $F(z)$ we found $q=2.03\pm0.08 $ A m,  in  good agreement  with previous estimates \cite{olegNPHYS2010}. For a rod of radius $r=d/2$, $q=M_{s}\pi d^{2}/4$  \cite{olegNPHYS2010}, using this result we computed the saturation magnetization as $M_{s}=(1.21\pm 0.03)\times 10^{6}$ A $m^{-1}$ which is in agreement with the available data of the magnetization of Neodymium rods, validating the dumbbell approximation when the rods are separated by a distance $z>d$, as previously pointed out \cite{Vokoun201155,Vokoun20093758,boyerAMJ1988}. 
 
 \section*{S2. Moment of Inertia, Static friction and Damping coefficient.}
The moment of inertia of magnetic rods around their local rotation axis is $I=M L^{2}/12$, where $M$ is the rod mass (assuming uniform density) and L its length. The extra moment of inertia given by the plastic holder of mass m, is negligible as it is short and light $m \sim 0.05\times 10^{-3}$ Kg. The mass of the magnetic rods used was measured individually for $50$ random rods which yielded: $M=(0.278\pm 0.003) \times 10^{-3}$ Kg. From this data we computed $I_{0}=8.41\times 10^{-9}$ Kg $m^{2}$. This value is used as input in the numerical simulations.

\begin{figure}[h]
\centering
\includegraphics[width=0.5\textwidth]{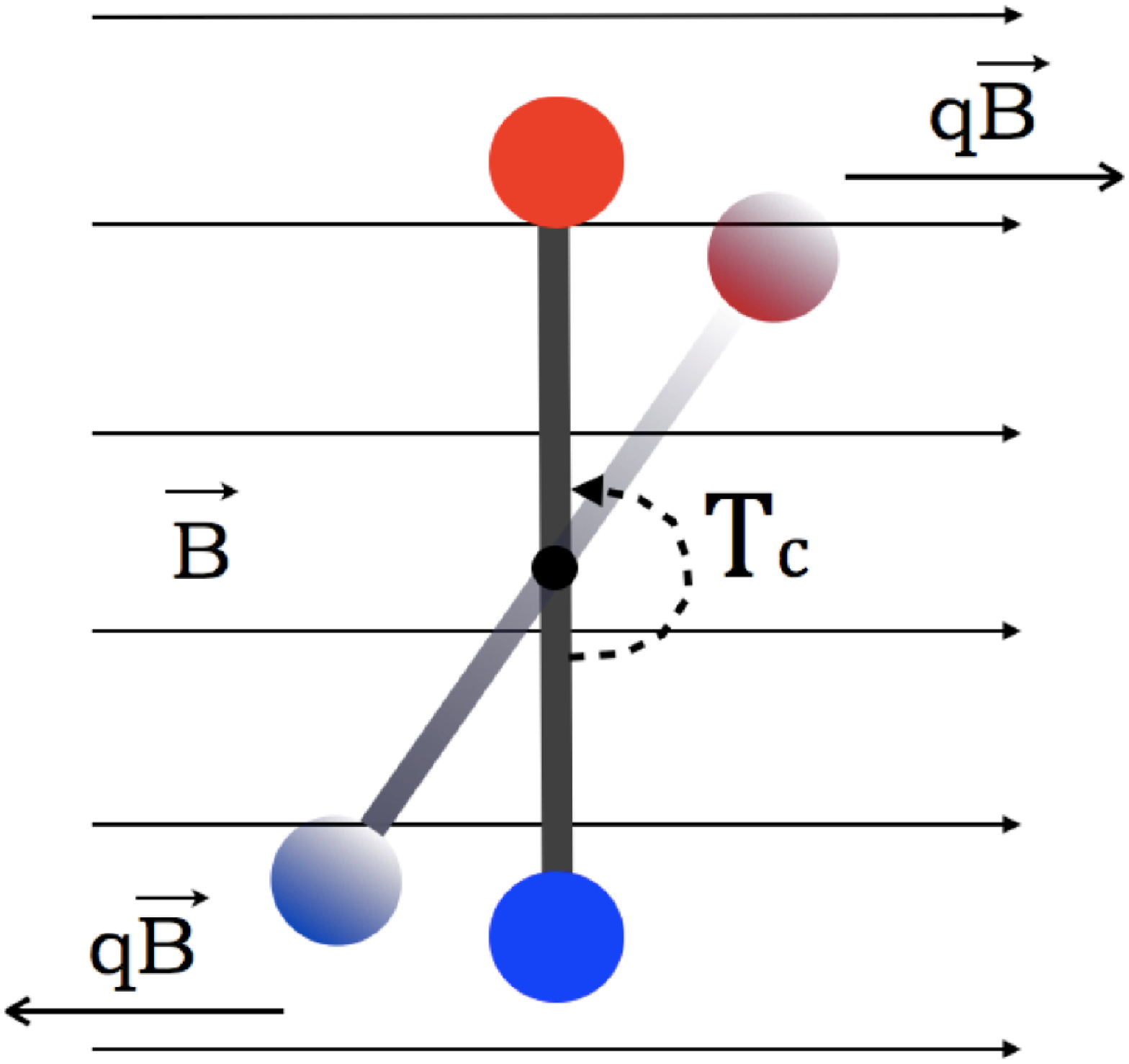}
\caption{\label{FIGS2} \textbf{Schematic of the setup used to measure static friction.} A uniform magnetic field is applied to a single rod producing a torque in each of its poles. Once the rod depart from its equilibrium position we recorded the value of the magnetic field and computed $T_{c}$.}
\end{figure}

We quantified the static friction on each rotor, by placing a single rod at the center of a Helmholtz coil, as illustrated in Figure \ref{FIGS2} and  measured the critical field at which the rod deviates from its initial position to find $B_{c} = (2.4 \pm 0.1)\times 10^{-4}$ T. Using $B_{c}$ we obtained  the critical torque for which the rod starts to move and found that $T_{c}=2a q B_{c}\sim 0.93 \times 10^{-5}$ N m. 

\begin{figure}[h]
\centering
\includegraphics[width=0.7\textwidth]{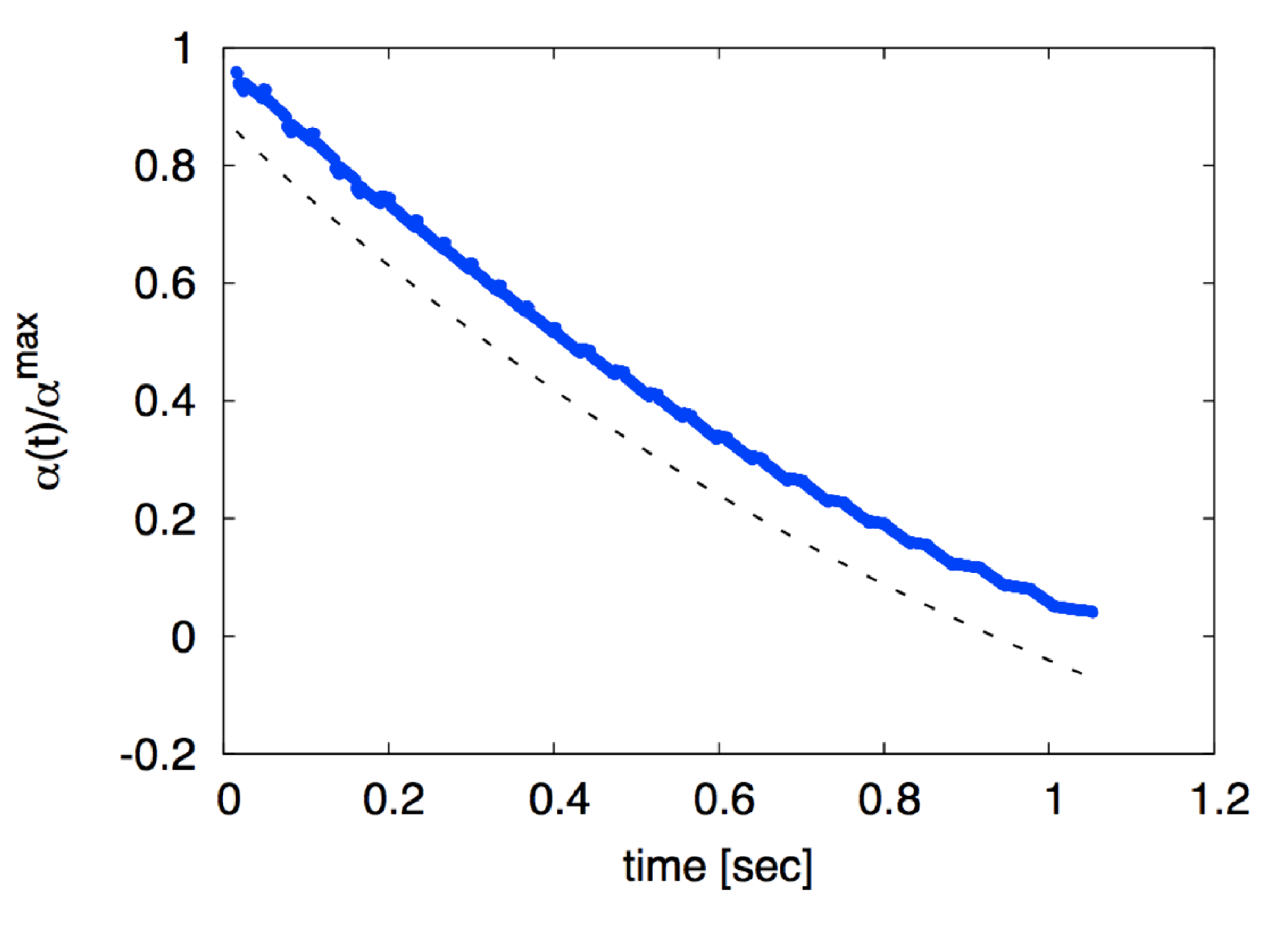}
\caption{\label{FIGS3} \textbf{Single rod dynamics.} In blue are the experimental data taken from standard imaging techniques and the dashed black is the best fit obtained using an exponential damping model, offset by 0.1 for clarity purposes.}
\end{figure}

To compute the damping coefficient $\eta$,  we isolated a single rotor and impulsively applied a torque to it, and then recorded its relaxation dynamics using a Phantom V$7.3$ high speed camera with frame rates between $1000$ fps (frames per second) up to $4000$ fps. Using standard imaging techniques we extracted  the evolution of $\alpha(t)$, which corresponds to a damped dynamics in absence of external forcing.  The damping is  computed directly by fitting it to  the  solution $\alpha\sim\exp(-t/\tau_{D})$, (Fig. \ref{FIGS3}) and thus we estimated the damping time of a single rod to be $\tau_{D}=0.83 \pm 0.18$ s.

\section*{S3. Single rod Dynamics and Triad}
A magnetic bar of length $L=2a$ and diameter $d \ll 2a$, behaves like two magnetic monopoles of equal strength but opposite sign located at each end of the bar \cite{Vokoun201155,Vokoun20093758}. The magnetic moment of each magnet is then given by $|\mathbf{m}|=2a q$, where $2a\mathbf{\hat{r}}$ is the vector separating the two poles and $\mathbf{m}$ points from S to N, so that poles of two distinct rods interact through Coulomb's law\cite{Vokoun201155,Vokoun20093758}.  The dynamics of a single rotor can be understood by examining its response to a uniform magnetic field along the $\hat{z}$ direction, with an equation of motion given by:
\begin{eqnarray}
I\frac{d^2\alpha}{dt^2}=T_{e}-\eta \frac{d\alpha}{dt}
\label{rod1}
\end{eqnarray}
where $I$ is the rotational moment of inertia of the rod (Supplementary  S2), $\eta$ is the damping coefficient of the hinge, and $T_{e}=2aqB_{0}\sin\left(\alpha\right)$ is the torque due to the action of an external magnetic field on the localized charges at the end of the rod. This simplified system has two natural time-scales: an inertial time $\tau_{B}=2\pi (I/2 a |q B_{0}|)^{1/2} \sim 0.003/|B_{0}|^{1/2}$ and a frictional time $\tau_{D}=I/\eta\sim 1 s$. Since $\tau_B = \tau_D$ for a critical field $B_{0}^*\sim 10^{-6}$ T, the dynamics of a single rod is underdamped in all our experiments.

\begin{figure}[h]
\centering
\includegraphics[width=0.7\textwidth]{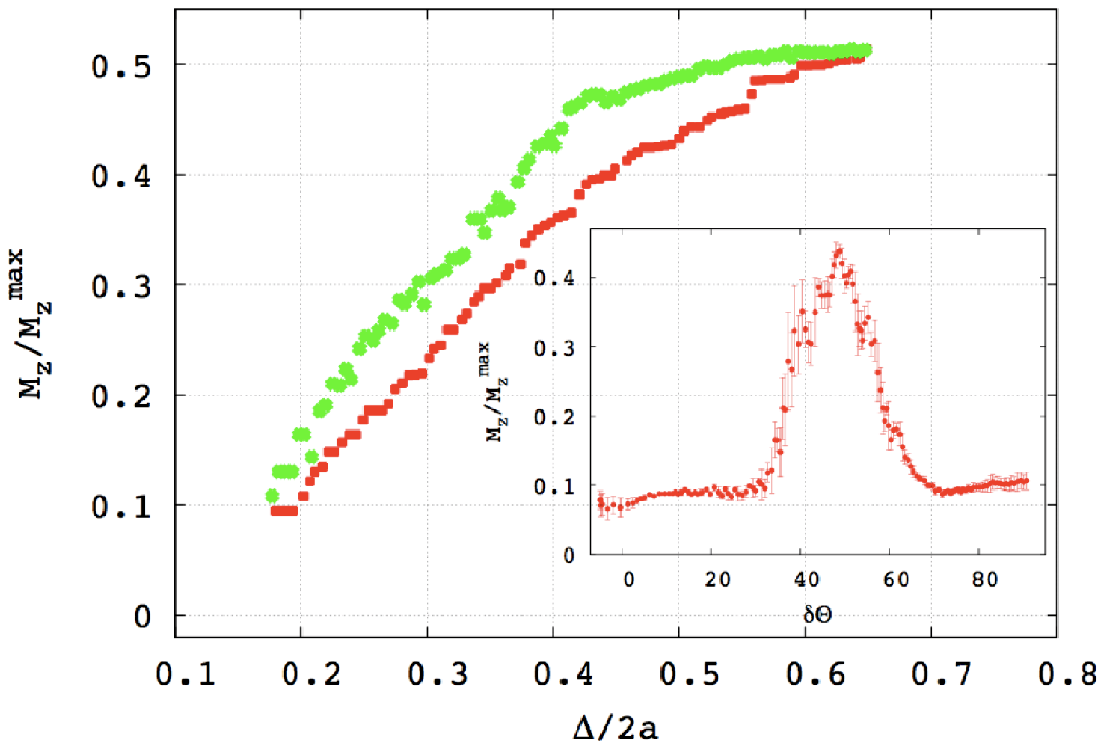}
\caption{\label{FIGS4} \textbf{$\Delta$ and $\theta$ dependence of the normalized magnetization for three rods in a $120^{\circ}$ star configuration.}  $M_{z}$ grows with $\Delta/2a$ and when the operation is reversed, the system shows hysteresis. The inset shows $M_{z}$ growing when one of the rotors change its azhimutal orientation respect to the others by $\delta \theta$}
\end{figure}

The minimal model where Coulomb interactions produces frustration in our system arises in the unit cell of three magnetic rods of length $L$ at $120^{\circ}$, that is frustrated in the plane. When $\Delta/L < 0.2$ the rods lie in the $x-y$ plane in a spin ice configuration as shown in Figure \ref{FIGS4}; as $\Delta/L$ is increased, the out of plane magnetization of the triad $M_{z}$ increases (Fig. \ref{FIGS4}), while when we reverse this operation, hysteresis is observed, a consequence of static friction in the system (see Fig. \ref{FIGS11}a for $M_{z}$ versus $\Delta/L$ in the lattice). We have used a single triad to examine the role of geometrical disorder by measuring $M_{z}$  when one of the hinges was rotated in the plane by $\delta\theta$ (Figure \ref{FIGS4} inset). We observe that  the system remains in the $x-y$ plane for small $\delta\theta$ ; however, when $\delta \theta\sim 35^{\circ}$ $M_{z}$ starts to increase rapidly until $\delta \theta\sim 50^{\circ}$ when one magnet becomes perpendicular to the plane.

\begin{figure}[h]
\centering
\includegraphics[width=0.7\textwidth]{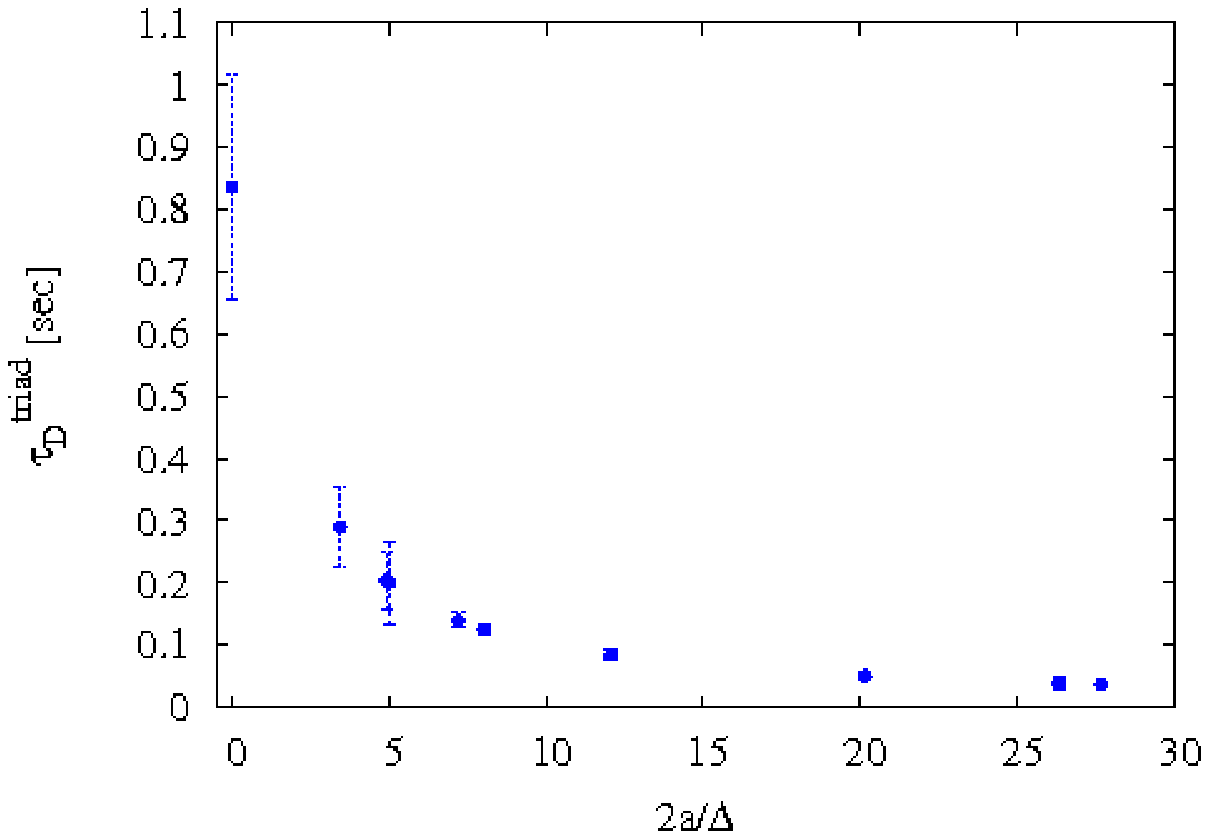}
\caption{\label{FIGS5} $\mathbf{\tau_{D}^{\rm{triad}}}$  \textbf{versus} $\mathbf{2a/\Delta}$.  The relaxation time of a single rotor in a triad configuration with two nearest neighbors.  $\tau_{D}^{\rm{triad}}\sim t_{rel}$, when $2a/\Delta<0.2$. The data $2a/\Delta=0$ is the damping time of a free rod, $\tau_{D}$.}
\end{figure}

\begin{figure}[h]
\centering
\includegraphics[width=0.7\textwidth]{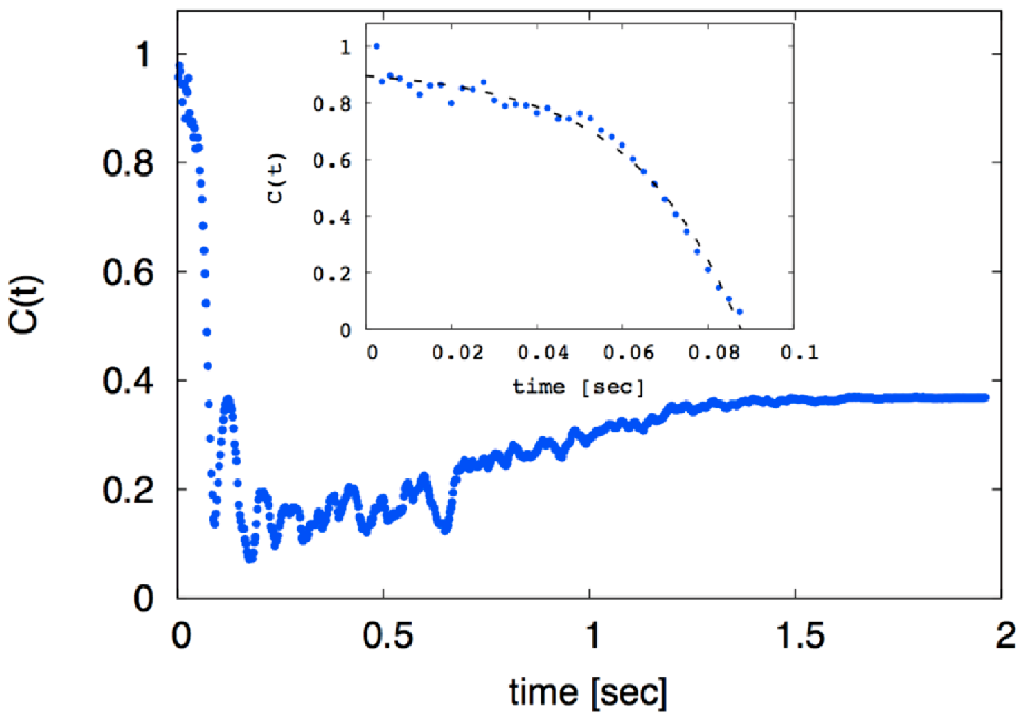}
\caption{\label{FIGS6}\textbf{Single particle autocorrelation, C(t).}   The (blue) points were obtained by computing the experimental single-particle autocorrelation function $C(t)=(\langle m_{i}(t)m_{i}(0)\rangle-\langle m_{i}\rangle^{2})/(\langle m_{i}^{2}\rangle-\langle m_{i}\rangle^{2})$ averaged over all rotors. Inset:  $C(t)$ for the first 0.08 seconds of lattice relaxation (blue) compared with the function $\cos(\exp(-t/t_{rel}))$ which is the the best fit for C(t) (dashed black curve).  From the fit we obtained the relaxation time of a single rotor in the lattice $t_{rel}=0.034$ sec.}
\end{figure}

The relaxation dynamics of the lattice made of unit cells is strongly dependent on the strength of the interactions and thus on the lattice parameters. For example, the damping time $\tau_{D}^{\rm{triad}}$ of one of the rotors belonging to a triad   grows with $\Delta$ (Fig. \ref{FIGS5}). When the $\Delta/2a$ approaches the lattice parameters, the dynamics of one rod in the triad approaches that of a rod in the lattice. Indeed when $\Delta/2a< 0.2$, we found $\tau_{D}^{\rm{triad}}\sim 0.04$ s, close to the relaxation time of a single rotor in the lattice, $t_{\rm{rel}}\sim 0.034$ s, which was found by computing the experimental single-particle autocorrelation function $C(t)=(\langle m_{i}(t)m_{i}(0)\rangle-\langle m_{i}\rangle^{2})/(\langle m_{i}^{2}\rangle-\langle m_{i}\rangle^{2})$ averaged over all rotors, and fitting its first 0.08 seconds of evolution to an  exponential decay as shown in Figure \ref{FIGS6}.

\section*{S4. Experimental Lattice}
All 352 rods were made out of Neodymium (NdFeB plated with NiCuNi), with saturation magnetization $M_{s}=1.2\times 10^{6}$ $A$ $m^{-1}$, length $L=2a=1.9\times 10^{-2}$ m, diameter $d=1.5\times 10^{-3}$ m and mass $M=0.28 \times 10^{-3}$ Kg. They were hinged at the plane that is equidistant from their N and
S magnetic poles, such that their axis of rotation crosses their center of mass. The hinges were made out of Acrylic-based photopolymer FullCure720 (Transparent) using a 3D printer Connex500 from ObJet Geometries. The hinges supporting the rods were introduced at the holes of an acrylic plate of size $0.61 \times 0.61$ m, defining the sites of a kagome lattice with lattice constant $l=\sqrt{3}(a+\Delta)$ where $\Delta = 4.675\times 10^{-3}$ m.  There was a small amount of geometrical disorder in the azimuthal orientation of the rotors, $\theta$,  due to lattice imperfections $\delta \theta_{\rm{max}}\sim 2^{o}$. We measured the azimuthal deviations of the rotors in the lattice when they were in the planar ($x-y$) configuration from lattice pictures (see below) to find that they follow a Gaussian distribution with mean $\bar{\delta\theta}=1.2^{\circ}$

The number of rods, and their positions in the lattice were unchanged between experiments otherwise mentioned. To examine the configurations of the system, we took high-resolution pictures out of ten different experimental realizations, and used standard imaging techniques to obtain $2D$ maps with the positions of each magnetic pole (South poles colored black). Each realization was accomplished after perturbing the lattice with an external dipole in order to produce a different initial and final magnetic state. In every occasion all vertices satisfied the ice rule. Away from the edges of the system, the spin ice rule was very robust, though small out of plane deviations larger than $\delta\alpha$, were sometimes observed at the edges of the lattice, a feature that becomes increasingly irrelevant for large systems.  The averaged value of the total normalized magnetization of the sample, along $x$ and $y$ directions typically achieved $\langle m_{x}\rangle \sim 0.12$ and $\langle m_{y}\rangle \sim -0.182$ respectively.
\subsection{Correlations}
In order to obtain the evolution of the nearest neighbor magnetic correlations in time, rotors were polarized using a Solenoid of 850 Turns with length 0.1 m and radius $r=0.4$ m made out of copper wire of diameter $d=0.51$ mm. The lattice was located inside the solenoid at its medial plane so that the radial component of the field was zero. The coil was connected in series with an ammeter and a regulated dc power supply serving as a current source.  We used currents up to
$i=3.5$ A to create a magnetic field strong enough to polarize our sample. At $i=1.5$ A, ($B_{z}\sim 3.2\times 10^{-3}$ T), all the N poles of the rods were pointing along the $\hat{z}$ direction. Next we turned the field off and waited until the rotors relaxed, simultaneously recording the evolution of the rods with a high speed Phantom V9.0 camera. We then analyzed these data using standard imaging techniques and extracted the full time trajectory, $\alpha_{i}(t)$, of each rod $i$ from movies like the one shown in Supplementary SM1. 

The three dimensional evolution of the rods is characterized by the vector
\begin{eqnarray*}
\vec{m}_{i}(t)=\left(\cos \theta_{i}\sin \alpha_{i}(t),\sin \theta_{i}\sin \alpha_{i}(t),\cos \alpha_{i}(t)\right) .
\end{eqnarray*}
thus the evolution of nearest neighbor rods correlations reads $S_{\alpha\beta}(t)\langle\vec{m}_{i}\cdot \vec{m}_{i+1}\rangle(t)=(1/n)\sum_{i}\vec{m}_{i}(t)\cdot\vec{m}_{i+1}(t)/|\vec{m}_{i}(t)\cdot\vec{m}_{i+1}(t)|$. A coarse grained charge at vertex k, at any time t, can be defined by $Q_{k}(t)=\sum_{i=1}^{3}q_{i}\cos \alpha_{i}(t)$. First, second and third nearest neighbors magnetic correlation as well as nearest neighbor charge correlations after relaxation are summarized in Figure \ref{FIGS8}.

\subsection{Lattice dynamics}
The fastest time scale of lattice relaxation is dominated by Coulomb interactions at the very onset of motion where all rods are in a configuration with their $\bf{S}$ poles  pointing along the $\hat{z}$ direction. During the first 0.07 seconds, two sublattices reach the plane before the remaining sublattice and organize in head to tail chains showing trend to ferromagnetic order. The experimental nearest-neighbor correlation decays exponentially,  with a characteristic Coulomb time scale $t_{c}=0.02$ seconds. To understand this, we  compute $\alpha(t)$ for a single rod using equation \ref{rod1}, where torques are due to Coulomb interactions with neighbors oriented at $120^{\circ}$ relative to each other in the chains along the $\hat{y}$ direction. The numerical solution of this minimal model allows us to find that $\cos(\alpha(t))$ decays exponentially with characteristic time equal to $t_{c}$. At leading order in $\alpha$, it also gave us a good estimate for $t_{c}= \frac{2^{3/4}}{q}\sqrt{4\pi I a/\mu_{0}}=0.02$ s. 

\begin{figure}[h]
\centering
\includegraphics[width=0.7\textwidth]{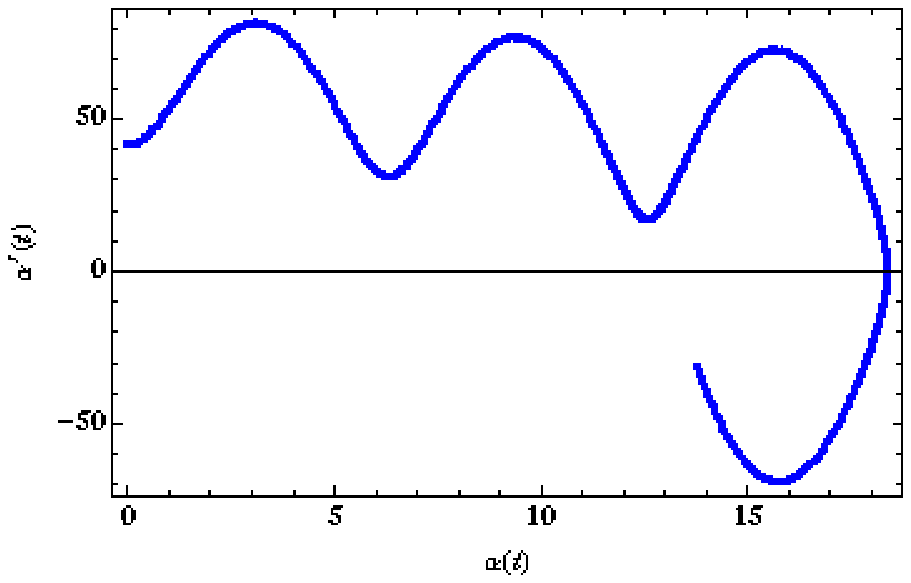}
\caption{\label{FIGS7} \textbf{Lattice dynamic at stage II.} The phase space trajectory corresponding to the solution of equation \ref{rotate} for a single rotor interacting with its four neighbors, after 0.45 seconds have elapsed.}
\end{figure}

\begin{figure}[h]
\centering
\includegraphics[width=0.7 \textwidth]{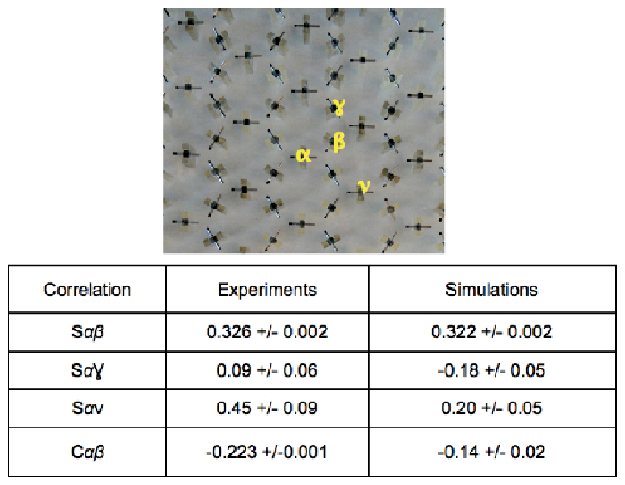}
\caption{\label{FIGS8}  \textbf{Correlations after relaxation.} \textbf{Top:} The honeycomb structure formed by connecting the spins of the kagome lattice. Each bar
element represents a rod oriented along the  axis. The Greek symbols label spins for  correlation
calculations. \textbf{Bottom:} First, second and third nearest neighbors magnetic correlation and nearest neighbors charge correlations for the experimental and numerical lattices. }
\end{figure}

Once all the rods reach the $x-y$ plane, the next relaxation stage involves the rotation of the rods around their center of mass. The amount of time rotations last in the system can be found by numerically solving the nonlinear differential equation of a damped rotor which rotate due to the torque generated by Coulomb interactions with its four nearest neighbors located at $120^{\circ}$ of each other in a mean field approximation:
\begin{eqnarray}
\alpha ''(t)=\Omega_{b}^2 \sin (\alpha
   (t))-\Omega_{\eta} \alpha '(t)
\label{rotate}
\end{eqnarray}
, with initial conditions $\alpha(0)=0$, $\alpha '(0)=V$, where $V$ is obtained from energy conservation at the onset of the rotor dynamics, $\Omega_{b}=\sqrt{2qa\langle B\rangle_{rms}}$, $\Omega_{\eta}=\eta/I \sim 1$ seconds and $\langle B\rangle_{rms}$ is the internal magnetic field due to the Coulomb interaction with its neighbors averaged over a cycle. For our experimental parameters, the phase space trajectory changes from open orbits into a dissipative attractor after 0.45 seconds: during this time the rotors  average about 4 full rotations before they begin to oscillate as shown in Figure  \ref{FIGS7}. As expected, this simplified model does not reproduce all details of the collective dynamics. However, it captures well the typical time scales for full librations and oscillatory behavior expected before the collective ground state is reached. To resolve the effect of collective dynamics we solved the equations of damped motion for each rotor Coulomb interacting with all the other rotors in the system using molecular dynamics simulations as described in section S5. 

\subsection{Phase diagram}
\begin{figure}[h]
\centering
\includegraphics[width=0.9 \textwidth]{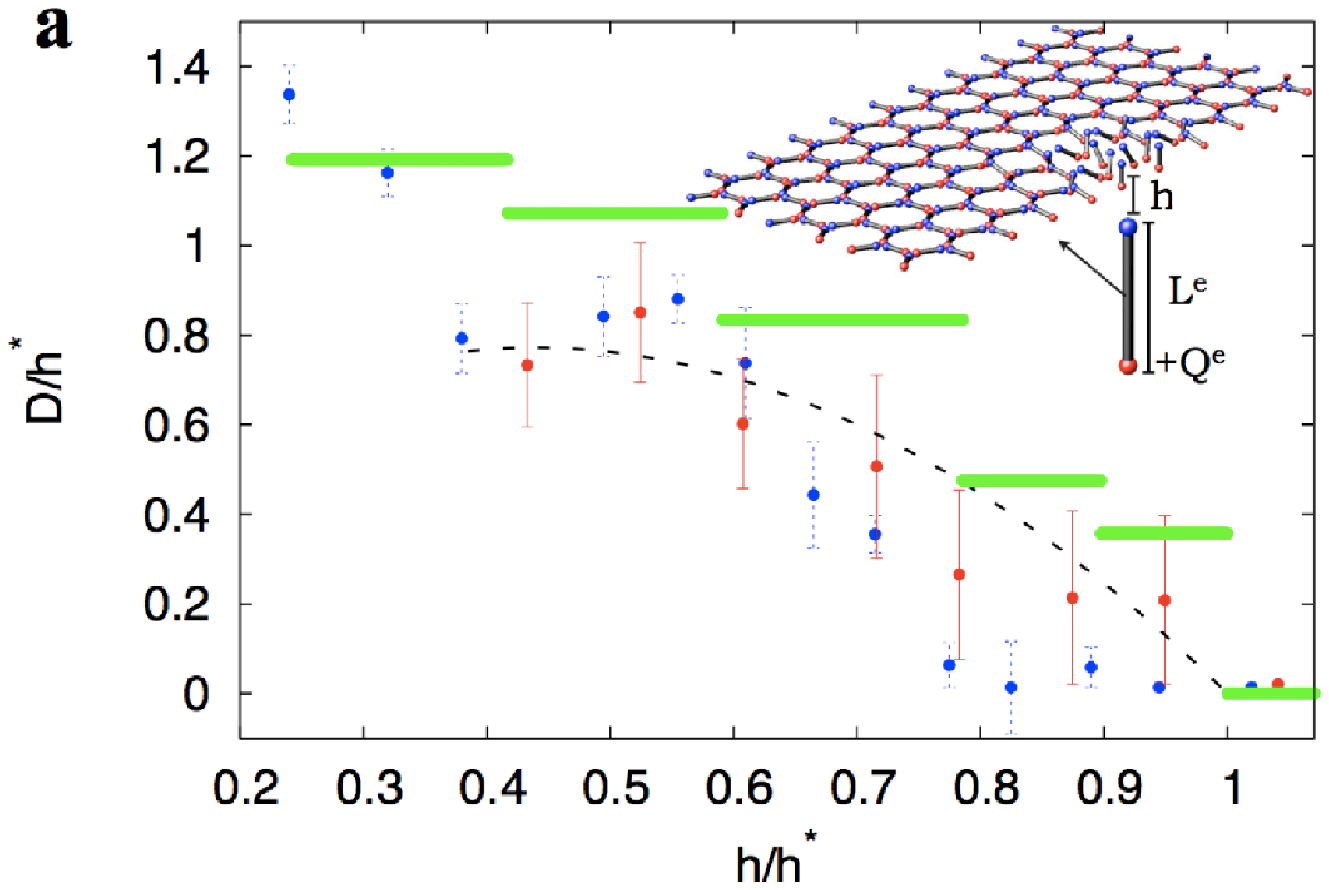}\\
\includegraphics[width=0.7 \textwidth]{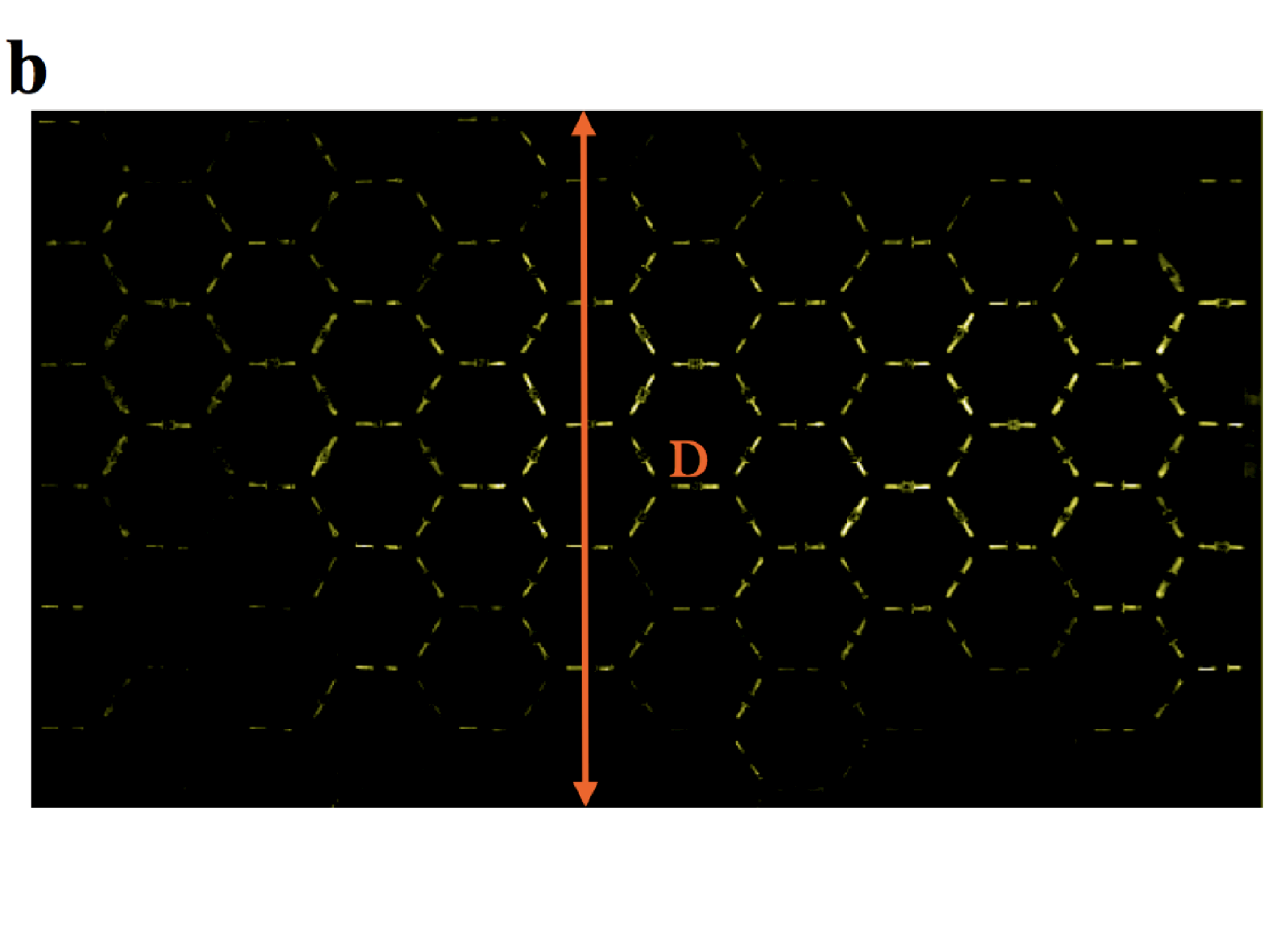}
\caption{\label{FIGS9} \textbf{ $D/h^{*}$ v/s $h/h^{*}$.} \textbf{a,} Experimental data showing the width $D$ of the stripe of the sample which is excited due to two dipoles of charges $Q^{e}=16q$ (blue) and $64q$ (red), in good agreement with the theoretical scaling (black dots) and the results from simulations (green) for $Q^{e}=64q$. The inset illustrates the experiment, where a dipole of length $L^{e}$ and charge $Q^{e}$ located at distance $h$ from the lattice is exciting a stripe of magnetic rods. \textbf{b,} Superimposed images obtained by computing the intensity difference between frames when the lattice is excited by an external dipole and a typical image from where we obtained experimental data for $D$. }
\end{figure}

To characterize the response of the system to external perturbations, we built a cart that can carry dipoles of length $L^{e}=0.135$ m, cross section radius $r^{e}={0.00635,0.00318, 0.00159}$ m and charges $Q^{e}=64q,16q$ and $8q$ respectively. This cart moves underneath the lattice with speed $v$ between $1-7$ m/s, set by the initial impulse provided to the device (inset Fig. \ref{FIGS9}a). The second experimental variable was the distance between the closest point of the dipole to the lattice, $h$, which varied between $0.42$ to $0.05$ m. In each realization we start with a different magnetic configuration where rotors lie at the $x-y$ plane at not apparent magnetic order. The magnetic configuration is a result of agitating the relaxed lattice moving an external dipole randomly underneath the lattice. Then from the middle of the left boundary of the sample, a dipole with magnetic charge $Q^{e}$ and length $L^{e}$ and of strength $B_{ext} \sim Q^{e}/h^{2}$, crossed the sample at a speed $v$ from one lattice edge to the other.
In order to distinguish between inertial and interaction regime, we quantified the response of the system computing the root mean square fluctuation (RMSF) in $\alpha$, about the rod's $x-y$ equilibrium position. This Lindemann like index $\delta_{Li}=\sum_{i,j}\frac{\sqrt{|\langle \mathbf{m_{i}}\mathbf{m_{j}}\rangle_{t}^{2}-\langle(\mathbf{m_{i}}\mathbf{m_{j}})^{2}\rangle_{t}|}}{|\langle\mathbf{m_{i}}\mathbf{m_{j}}\rangle_{t}|}$, quantifies the angular fluctuations in the context of the parameters of the system, where $\langle\rangle_{t}$ means the temporal average. The threshold value of the Lindermann parameters $\delta_{Li}^{th}=0.5$. When $\delta_{Li}<\delta_{Li}^{th}$, the system is in an inertial regime. The boundary between interaction and inertial regimes can be obtained using a single particle approximation by equating the impulse a
rotor feels when the external dipole is moving at speed $v$, at a distance $d(t)=\sqrt{h^{2}+(v t)^{2}}$, from it, to the
change in the angular momentum of the rotor: $\delta L=\int (F(t)\times a) dt$, where $F=\mu_{0}qQ^{e}/(h^{2}+(v t)^{2})$. Integrating over time in the right side between $-\beta/v$ and $\beta/v$, where $\beta=\sqrt{(h^{*2}h)^{2/3}-h^{2}}$, yields the change in the angular momentum of the rod due to its interaction with the upper charge of the dipole: $\delta L \sim \mu_{0}qQ^{e}a/(2\pi h I v)$ ($h^{*}$ is the threshold height for which the Coulomb interaction between a rod and the external dipole overcomes the static friction, as explained in the paper). For an arbitrary size of oscillations, $\delta_{Li}^{th}$, we
can get an estimation of how $h$ change with $v$ by integrating over time at the left side of the previous
equation to get : $h\sim h^{*}(\mu_{0}Q^{e}qa/I)^{3/2}/v^{3}$, which gives good
account of experimental results.  The interplay between the ratio of internal to external magnetic forces and the speed of the dipole, determines in average the oscillations amplitude in a belt of the sample whose width $D$ can be computed in the quasi-static approximation (Fig. \ref{FIGS9}), in terms of local interactions, static friction, and interactions with the top charge of the external dipole yielding $D/h^{*}\sim \sqrt{(h/h^{*})^{2/3}-(h/h^{*})^{2}}$, when $h<h^{*}$.

\section*{S5. Molecular Dynamics simulation}
\begin{figure}[h]
\centering
\includegraphics[width=0.7 \textwidth]{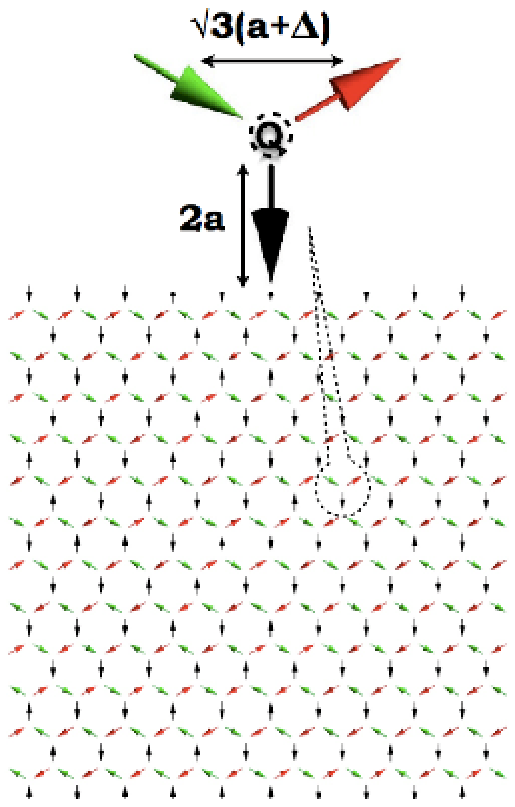}
\caption{\label{FIGS10} \textbf{Numerical lattice:} Rods in a  typical configuration fulfilling spin-ice
rules after the numerical lattice has reached relaxation.}
\end{figure}

Numerical results were obtained by direct numerical integration of the equations of motion for each rod interacting with all the others  351 rotors via Coulomb interactions in the dumbbell picture using a Verlet method with an integration time step $\Delta t=10^{-4}$ at zero temperature. In all simulations experimentally measured parameters for lattice constant, damping, inertia  and charge of the rotors were used. Equilibration was accomplished after 2 seconds of real time lattice evolution ($2\times 10^{4}$ time steps). To check the system size dependence on the calculated quantities, we compared our results with those for a system with $n=900$ rods. As there was no significant difference in the results between
the $n=900$ and $n=352$ systems, we used the later number for all our simulations. 

\begin{eqnarray}
I\frac{d^2\alpha^{(i)}}{dt^2}=T_{e}^{(i)}-\eta \frac{d\alpha^{(i)}}{dt}  \mbox{\hspace{0.5cm},where: }  \\ \nonumber
T_{e}^{(i)}=\left(\frac{a \mu_{0}}{4\pi}\right)\sum_{j\neq i}q_{i}q_{j}\hat{r}_{i}^{cm}\frac{\vec{r}_{ij}}{|\vec{r}_{ij}|^{3}}
\label{rod2}
\end{eqnarray}
$\vec{r}_{ij}$ is the vector joining  charges $q_{i}$ and $q_{j}$, and $\hat{r}_{i}^{cm}$ is the unit vector pointing from the center of mass (cm) of the rod where
charge $q_{i}$ belongs.
We find that the numerical lattice always fulfills the spin ice rules as shown in Figure \ref{FIGS10}. Correlations after equilibration compared with the experimental values (Table of Fig. \ref{FIGS8}), show that nearest and second neighbor correlations compare well while third neighbors and charge correlations are smaller than in the experimental counterpart. We attribute this to the absence of static friction in the numerical model. To complement the experimental phase diagram we performed simulations using a point-like magnetic charge $Q^{e}=64q$, moving in a slightly larger parameter space that the one experimentally explored , $\Sigma=(0.005,10)\times (0.05,0.5)$. Our numerical constructed  findings in the inertial and interaction regimes fall inside the experimental one, showing that the basics physics is captured well by the numerical model.

\begin{figure}[h]
\begin{center}
\includegraphics[width=0.7\textwidth]{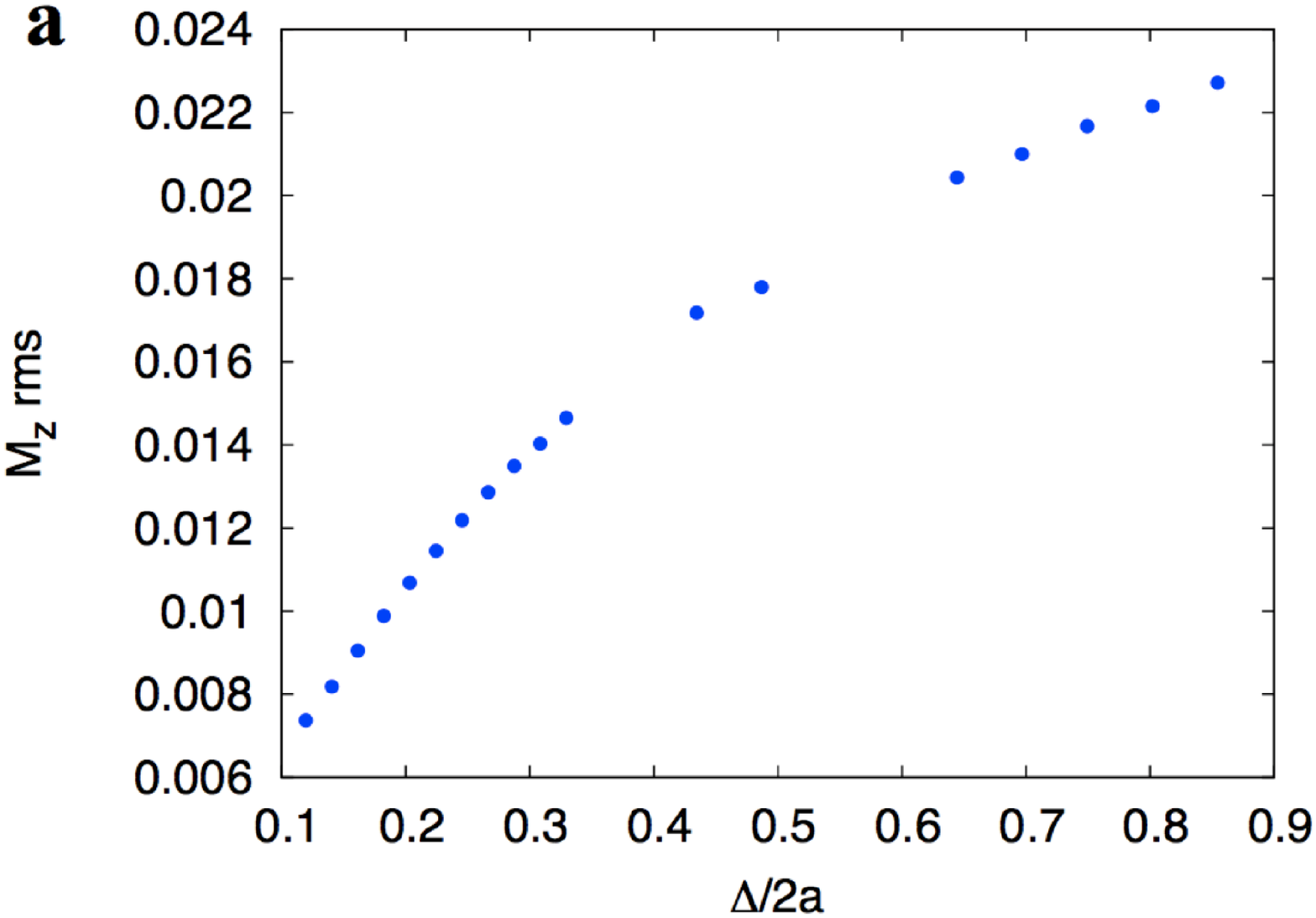}\\
\includegraphics[width=0.7\textwidth]{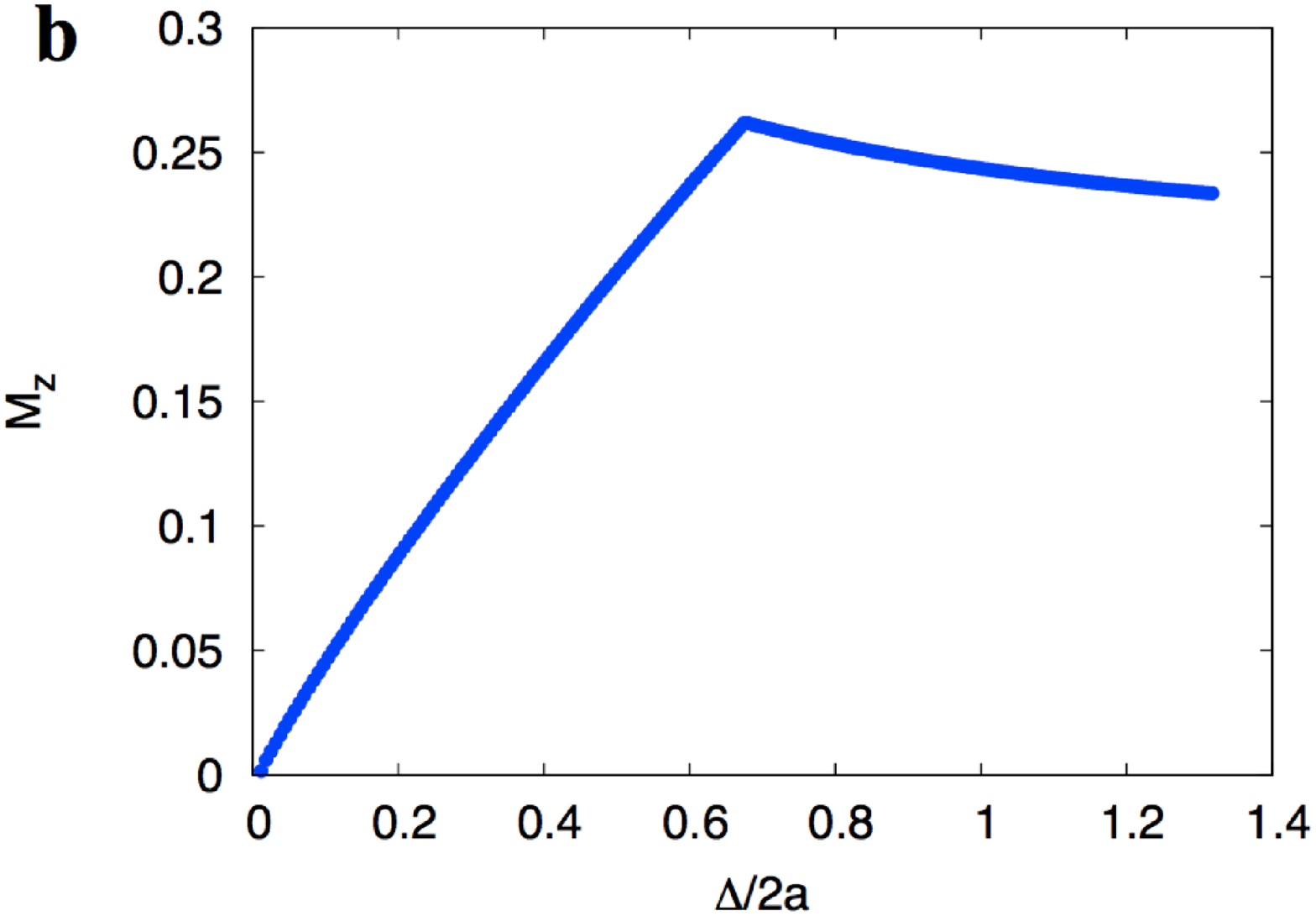}
\caption{\label{FIGS11} \textbf{Numerical $M_{z}$ v/s $\Delta$.}\textbf{a},  RMS deviations out of the $x-y$ plane averaged for all the n rods of the lattice $M_{z}rms$  v/s  $\Delta/2a$  after 4 seconds of simulation have elapsed. As the distance between nearest poles increases, rotors are more prone to choose positions which depart from the $x-y$ plane at equilibrium.
 \textbf{b,}  Average magnetization along the $z$ direction of a triad of rotors  v/s $\Delta/2a$.   $M_{z}$ was obtained computing the Coulomb interaction between the six poles and minimizing numerically the energy en the domain $(0,2\pi)^{3}$.}
\end{center}
\end{figure}

Molecular dynamics simulations also permitted the examination of the out of plane lattice fluctuations versus variations in $\Delta/L$. Figure \ref{FIGS11}a shows  the lattice RMS deviations out of the $x-y$ plane averaged for all the n rods of the lattice $M_{z}\rm{rms}=\sqrt{\sum_{i}(\cos{\alpha_{i}}^{2})}/n$ growing linearly for $\Delta < 0.3L$. They show a behavior qualitatively similar to the total $M_{z}$ of three dumbbells interacting via Coulomb, in a star configuration, obtained from energy minimization,  Figure \ref{FIGS11}b for $\Delta < 0.4L$. An extended numerical study will be published elsewhere. 

\section*{S6. Theory: Dipoles v/s Dumbbells}
Once all rotors were placed in the kagome lattice, we found that the honeycomb spin ice phase is very robust: any time  a rotor located at the bulk was displaced with our finger tips,  it came back to the $x-y$ plane after a few oscillations. When we displaced it strongly so that a rotor flipped, locally new configurations in the spin ice manifold formed from sequential first-nearest-neighbor flips alone. Sequentially flipped second-nearest neighbor pairs were never observed, unless linked by a shared first nearest neighbor. Rotors being part of a closed magnetic loop were particularly stiff to external perturbations thus; single-particle spin-flipping suggests that flipping dynamics depends on local magnetic configurations. 

We examined two limit cases in order to understand these findings. In the first one, we let the magnets lie far apart ($\Delta/a > 0.4$) so that a Hamiltonian with nearest neighbors dipolar interactions is a good approximation \cite{Vokoun201155,Vokoun20093758,boyerAMJ1988}. We found that the dipoles having a continuous U(1) symmetry, along their local axis prefer configurations in the  $x-y$ plane in the honeycomb spin ice manifold. The ground state energy is degenerate, $E=-\frac{7D}{8}N_{t}$, where $N_{t}$ is the total number of triangles of the kagome lattice and $D\sim \mathit{O(a/\Delta)}\sim 10^{-5}$ J, sets the energy scale for interactions. The cost of raising a rod is $\delta E = \frac{7D}{2}$, decreasing as $\Delta$ increases. This model also provides an easy estimation of the local energy configurations for a triad of rotors shown in Figure \ref{FIGS1} where the ones satisfying the spin ice rule are energetically favorable since head to tail magnetic configurations are energetically favorable ($-1.75 D$ instead of $1.75 D$ of tail to tail or head to head configurations) which explains also why magnets taking part of closed loops are particularly stable to perturbations.

In the opposite limit, relevant for our experiments, rotors are brought close to each other such that $\Delta/a<0.4$, and thus a dumbbell model is a reasonable description \cite{Vokoun20093758}. The ground state of our system can be understood in terms of magnetic dumbbells with magnetic charges at the ends of each rod. For a single triad of dumbbell dipoles, each having a length $2a$ and a charge of magnitude $\pm q$ on each dipole head and tail respectively, the full Coulomb interaction between magnetic  charges contains $12$ terms and an arbitrary constant that represents the self energy of this system. The minimal distance between the origin and the closest charge is $\Delta$, therefore in the limit  $ \Delta \rightarrow 0$ the energy will be given only by the three  divergent Coulomb terms due to the interactions between the three charges $1,2,3$, that are closest to the origin. Thus, at leading order, the interaction energy is given by $U_{\Delta}\sim U_{12}+U_{23}+U_{31}$, where $U_{\Delta}$ is clearly divergent. In the same spirit as in QED and QCD \cite{PhysRevB.68.024508}, we describe the basic physics by paying attention only to the most divergent quantities and can understand the optimal charge configuration using the fact that $U_{\Delta}\sim g/\Delta$. Thus minimizing the Coulomb energy is equivalent to minimizing g. In the limit when $ \Delta \rightarrow 0$,  $a$ is the only length scale. Then the leading order term in g will only depend on the geometry, and any needed change in the sign of the charge product will appear as a $\pi$ change in the polar rotation angle. Therefore in considering the lowest energy configuration of three interacting dumbbells that are at a distance $\Delta$ from each other, we find that:
\begin{eqnarray*}
&g(\alpha_{1},\alpha_{2},\alpha_{3})=
\\ \nonumber&(5+\cos\alpha_{1} \cos \alpha_{2}
-\sin \alpha_{1}\left( -3+\sin \alpha_{2}\right)
-3\sin \alpha_{2}
)^{-1/2}+ \\ \nonumber
&(5+\cos \alpha_{1} \cos \alpha_{3}
+\sin \alpha_{1}\left( 3+\sin\alpha_{3}\right)
+3\sin\alpha_{3}
)^{-1/2}+\\ \nonumber
&(5-\cos\alpha_{2} \cos\alpha_{3}
-\sin\alpha_{2}\left( 3+\sin\alpha_{3}\right)
+3\sin\alpha_{3}
)^{-1/2}
\label{eq:h3}
\end{eqnarray*}
where $\alpha_{i}$ are the rotation angles measured relative to the vertical axis (polar angle). The global minimum for $g(\alpha_{1},\alpha_{2},\alpha_{3})$ occurs for planar configurations corresponding to the spin ice case, i.e. $(\alpha_{1},\alpha_{2},\alpha_{3})=\pm (\pi/2,\pi/2,-\pi/2)$ or permutations. Thus, the basic building blocks of our  experiment follow spin ice rules as the shortest distance between magnetic poles goes to zero asymptotically: the initial $U(1)$ symmetry for each rotor is reduced into a $Z_{2}$ Ising like symmetry.

\section*{Movies}

\section*{Caption Movie SM1}\textbf{Experimental lattice dynamics.} In the initial state all the S poles of the lattice n rotors are pointing along $\hat{z}$. At $t=0$, the magnetic field is switched off and the lattice relaxes over a period of $2$ seconds. This dynamical process is shown here along with the evolution of the nearest neighbor magnetic correlations. In the final state all the magnets lie in the $x-y$ plane in a honeycomb spin ice magnetic order.

\section*{Caption Movie SM2}\textbf{Computational lattice dynamics.} In the initial state all the S poles of the lattice n  rotors are pointing along $\hat{z}$. At $t=0$, the magnetic field is switched off and the lattice relaxes  over a period of $2$ seconds. This dynamical process is shown here along with the evolution of the nearest neighbor magnetic correlations. In the final state all the magnets lie in the $x-y$ plane in a honeycomb spin ice magnetic order.

\begin{figure}[h]
\begin{center}
\includegraphics[width=0.5\textwidth]{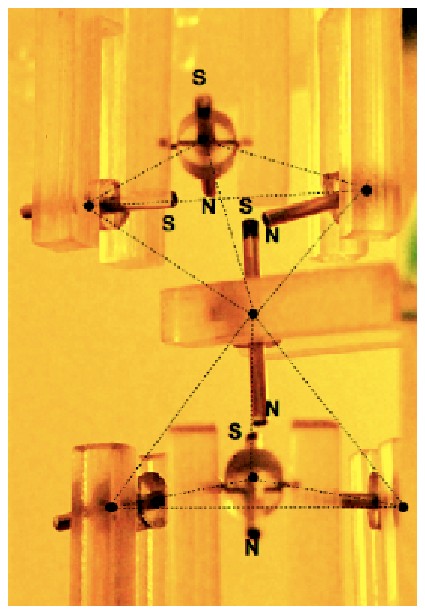}
\caption{\label{FIGS12} \textbf{A minimal 3-D generalization of our system.} A tetrahedral configuration like the one found in the Pyrochlore lattice was created placing three acrylic plates one on the top of the other, the bottom and top plates contain three rotors defining an equilateral triangle and the middle plate contains one rotor located equidistant from the others.}
\end{center}
\end{figure}